\documentclass{article}
\usepackage[margin=1in]{geometry}
\usepackage{amsmath, amsthm}
\usepackage{booktabs, subcaption, threeparttable}
\usepackage{caption}
\usepackage{comment}
\usepackage{xcolor, soul}
\usepackage{hyphenat}
\usepackage{graphicx}
\usepackage{pifont}
\usepackage{url}
\newcommand{\cmark}{\ding{51}}%
\newcommand{\xmark}{\ding{55}}%
\newtheorem{example}{Example}[section]
\theoremstyle{definition}
\newtheorem{definition}{Definition}

\newcommand{\revA}[1]{#1}

\newcommand{\revGeneral}[1]{#1}
\newcommand{\np}[1]{}
\newcommand{\rs}[1]{}
\newcommand{\gr}[1]{}

\begin{document}

\title{BEACON: Budget-Aware Entity Matching Across Domains\\
\large Extended Technical Report\footnote{This paper is the extended version of
``BEACON: Budget-Aware Entity Matching Across Domains,'' to appear in
\textit{Proc. ACM SIGMOD International Conference on Management of Data (SIGMOD 2026)}~\cite{pulsone2026beacon}.}}

\author{
Nicholas Pulsone \\
Worcester Polytechnic Institute \\
\texttt{nbpulsone@wpi.edu}
\and
Roee Shraga \\
Worcester Polytechnic Institute \\
\texttt{rshraga@wpi.edu}
\and
Gregory Goren \\
eBay Research \\
\texttt{grggoren@gmail.com}
}

\date{}

\maketitle

\begin{abstract}
  Entity Matching (EM)--the task of determining whether two data records refer to the same real-world entity--is a core task in data integration. Recent advances in deep learning have set a new standard for EM, particularly through fine-tuning Pretrained Language Models (PLMs) and, more recently, Large Language Models (LLMs). However, fine-tuning typically requires large amounts of labeled data, which are expensive and time-consuming to obtain. In the context of e-commerce matching, labeling scarcity varies widely across domains, raising the question of how to intelligently train accurate domain-specific EM models with limited labeled data. In this work we assume users have only a limited amount of labels for a specific target domain but have access to labeled data from other domains. We introduce BEACON, a distribution-aware, budget-aware framework for low-resource EM across domains. BEACON leverages the insight that embedding representations of pairwise candidate matches can guide the effective selection of out-of-domain samples under limited in-domain supervision. We conduct extensive experiments across multiple domain-partitioned datasets derived from established EM benchmarks, demonstrating that BEACON consistently outperforms state-of-the-art methods under different training budgets.
\end{abstract}

\section{Introduction} \label{sec:introduction}

Entity Matching (EM)--also known as Entity Resolution or Record Linkage--is a fundamental task in the data integration pipeline. At its core, EM involves determining whether two data records refer to the same real-world entity. It is widely applied in scenarios such as de-duplication, where the goal is to group samples in a single dataset into clusters of distinct entities. EM is a longstanding problem in the data management and data science communities and has been studied extensively over the past several decades~\cite{fellegi1969theory, hernandez1995merge, Bilenko03, bhattacharya2007collective, barlaug2021neural, awick2025lmsurvey}.

Historically, EM was addressed using string similarity metrics and probabilistic models~\cite{fellegi1969theory, winkler1990string}, followed by the development of rule-based systems~\cite{hernandez1995merge, monge1996field}. With the advent of learning methods, more flexible and adaptive models emerged~\cite{Bilenko03, bhattacharya2007collective}, ultimately leading to the adoption of deep learning methods based on neural networks~\cite{ebraheem2017deeper, mudgal2018deep}. While recent studies have employed Large Language Models (LLMs) for EM~\cite{peeters2023llm}, the fine-tuning of Pretrained Language Models (PLMs)--such as BERT and RoBERTa--on task-specific EM datasets~\cite{brunner2020entity, peeters2021dual} remains a common paradigm~\cite{awick2025lmsurvey}. This is largely because LLMs impose significant computational overhead during inference~\cite{zhang2025deep}--even in zero-shot settings--and incur high monetary costs when accessed via hosted APIs~\cite{peeters2023llm}. Among PLM-based approaches, DITTO~\cite{li2020ditto} represents the state-of-the-art (SOTA) EM framework. This work builds on the DITTO pipeline and focuses on optimizing EM in settings with limited training data.

Many real-world EM applications involve heterogeneous data spanning multiple \emph{domains}~\cite{primpeli2020profiling, fu2021hierarchical}. For example, the WDC benchmark~\cite{peeters2023wdc} partitions data by product category, and other standard EM datasets--such as Abt-Buy and Amazon–Google Products--exhibit similar domain structure~\cite{kopcke2010evaluation}. Consequently, it is often intuitive to train and evaluate separate models for each domain to optimize performance~\cite{shah2018neural}, to-be-defined in Section~\ref{sub:prob_def}. This setup enables a more informative evaluation protocol than traditional EM, as it reveals how model performance varies across domains.

Processing diverse domains often coincides with substantial variation in data characteristics, including the availability of labeled examples. As machine learning methods, particularly deep learning-based language models, have become increasingly prevalent in EM, there is a growing reliance on large labeled datasets. However, curating such datasets at scale is expensive and labor-intensive~\cite{genossar2023battleship,shraga2022humanal}. To capture this constraint, we consider an \emph{annotation budget}~\cite{sarawagi2002interactive, kasai2019low, genossar2023battleship}, which we define as a hard limit on the number of labeled samples available for training an EM model (e.g., 5,000 samples). This budget reflects the finite labeling resources available to a practitioner.

In realistic deployment settings, practitioners often possess labeled data for a single domain of interest, alongside abundant unlabeled data from related domains~\cite{settles2009active}. As a result, the primary bottleneck becomes the cost of acquiring additional labels across domains. In e-commerce for example, organizations may have extensive historical or cross-category product data but have annotated only the specific domain they seek to improve. Another common scenario is that an organization already has labeled data spanning multiple categories, but must support a new domain with limited or no labeled data and only a small additional annotation budget. In such cases, a fundamental question arises: how can we intelligently select which samples to label in order to train effective EM models under a fixed annotation budget? This paper aims to address this question by \emph{exploring how EM can be performed effectively across domains under a fixed annotation budget, and introducing \textbf{BEACON}---a novel budget-aware framework designed to address this challenge.}

To illustrate the motivation for our setting and approach, we now turn to a concrete example.

\begin{figure}
  \centering
  \includegraphics[width=0.75\textwidth]{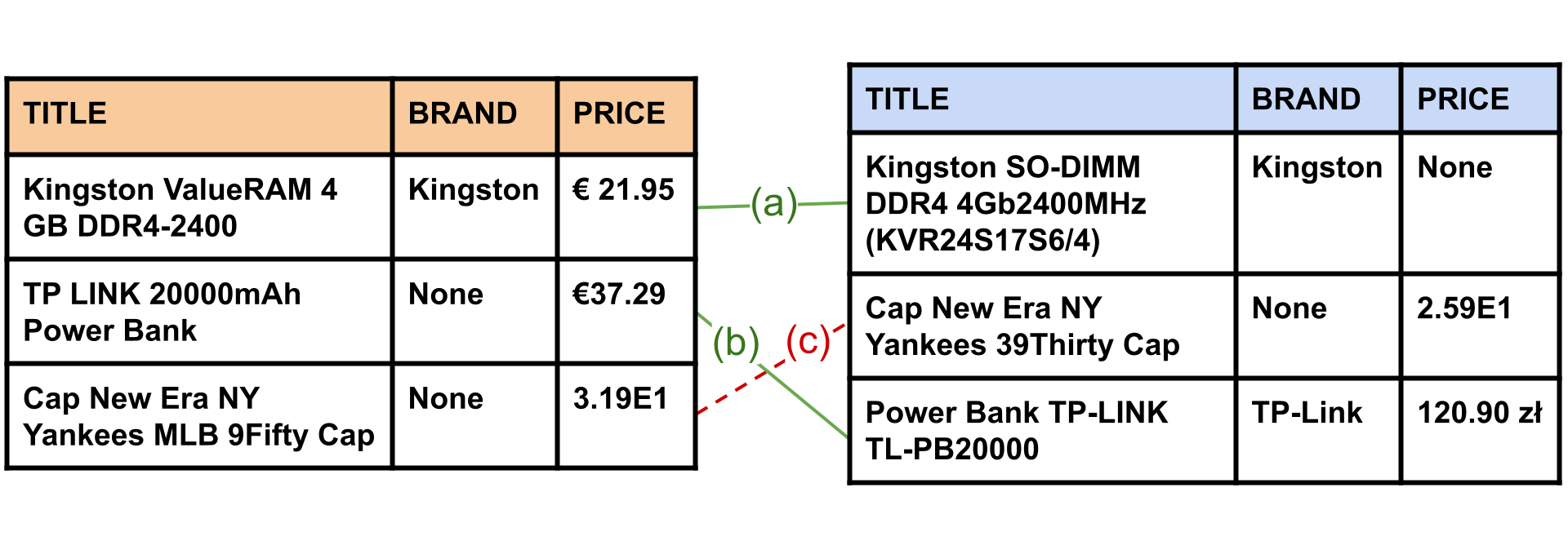}
  \caption{Performing EM between two e-commerce datasets. Each sample compares two product records and is derived from the Web Data Commons Multi-Dimensional Entity Matching Benchmark (WDC)~\cite{peeters2023wdc}.} 
  \label{fig:em_example}
\end{figure}

\begin{example} \label{ex:em_example}
A common EM process involves forming pairs of entities that could potentially refer to the same real-world object. This is typically done using a procedure known as \emph{blocking}~\cite{simonini2016blast, papadakis2020blocking, thirumuruganathan2021deepblockds}. Blocking considers all possible combinations of records and eliminates unlikely matches early in the pipeline using computationally inexpensive rules and heuristics. The result is a reduced search space that enables a more fine-grained matching procedure. For this example, we assume the output of blocking is already given. 

Figure~\ref{fig:em_example} shows three entity pairs produced by blocking, with each pair comparing two product records from e-commerce datasets. The samples span multiple product categories and illustrate the types of candidate pairs that serve as inputs to EM models. A solid green line between records indicates a match, while a red dotted line indicates a non-match. Samples (a) and (c) represent matches, where both entities in the pair refer to the same product (though this is not known during matching). For example, sample (a) compares the same RAM stick, and sample (c) compares the same power supply. Conversely, sample (b) is a non-match: although both products are baseball caps from the same brand, they are distinct items. The record data is incomplete--as can be seen with missing ``brand'' values--highlighting the challenges of real-world data systems and EM in practice.
\end{example}

\begin{figure}
  \centering
  \includegraphics[width=0.75\textwidth]{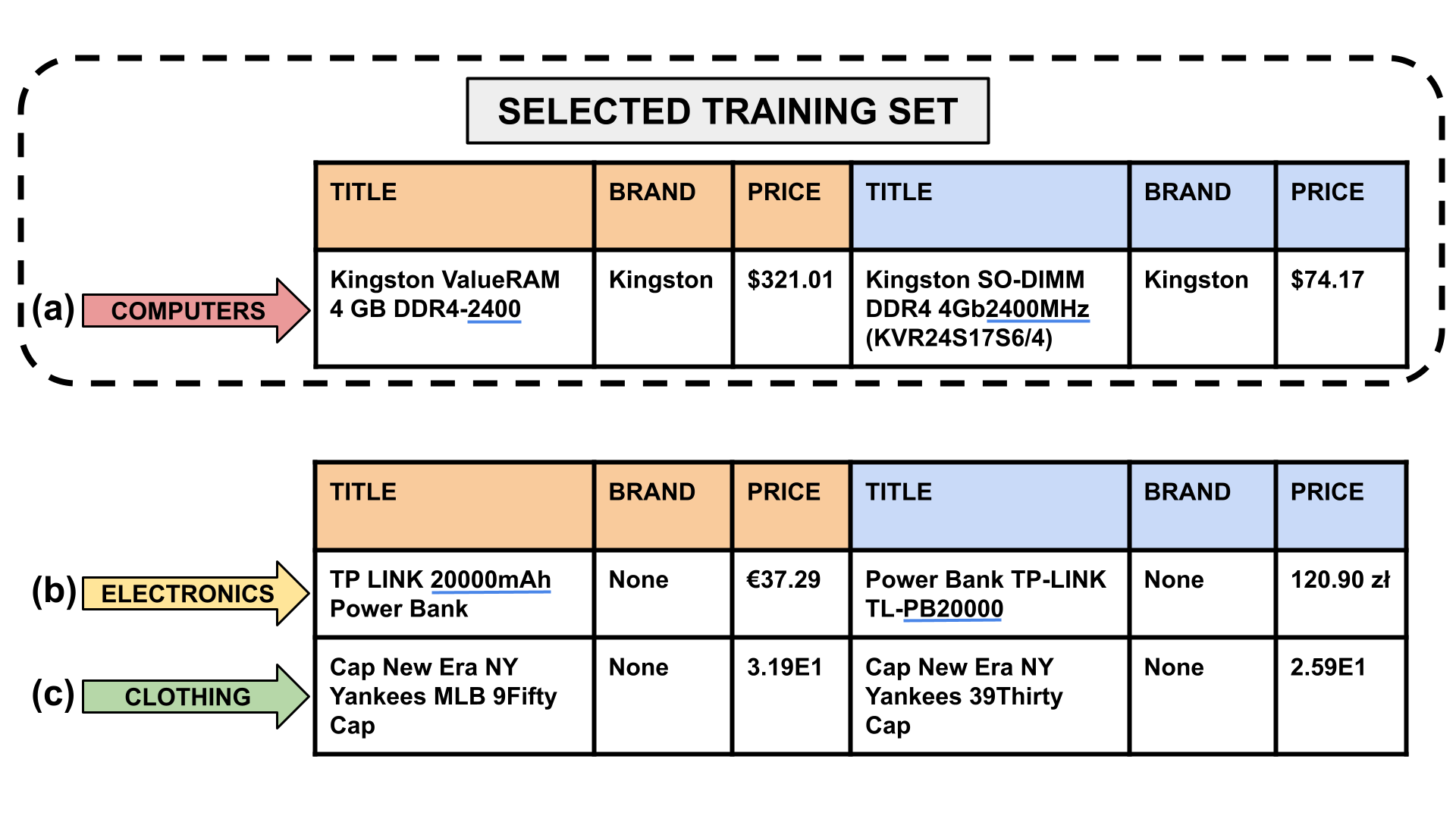}
  \caption{Entity matching samples from Figure~\ref{fig:em_example} grouped by the product category domain. The goal is to select a pairwise sample to complement the selected training set for EM in the ``Computers'' product category. }
  \label{fig:neighbor_example}
\end{figure}

Building on the previous example, we now examine how EM can be approached \emph{across domains}--at the level of individual product categories. Such category-level partitioning is common in e-commerce EM benchmarks~\cite{primpeli2020profiling, peeters2023wdc}, which explicitly organize data by product type to enable domain-specific evaluation. This approach requires a careful selection of samples to train an effective EM model, especially under a limited training budget.

\begin{example} \label{ex:neighbor}
Suppose we aim to train a model for the ``Computers'' category. Figure~\ref{fig:neighbor_example} shows the same three pairwise samples from Figure~\ref{fig:em_example}, now labeled by product category. Note that any pairs spanning different product categories are automatically removed during blocking, allowing the remaining data to be partitioned into distinct category-specific groups.
To build 
a training set for “Computers,” we first include all in-domain pairs, as they are most likely to be relevant to our model. In our toy example, only pair (a) belongs to the domain and is selected first. If room remains in the budget after including all in-domain samples, we 
need to decide which out-of-domain examples are most ``beneficial.'' 

Sample (b), from ``Electronics,'' might be a natural candidate. Intuitively, “Electronics” is ontologically closer to “Computers” than ``Clothing'' is, meaning that, in an abstract product taxonomy or to a human judge, the computers and electronics categories 
have the same semantic neighborhood, or at least more so than clothing does. More importantly however, both domains exhibit similar structures, 
such as 
numerical specifications that may vary in format or include/exclude units. Sample (c), from ``Clothing,'' may also be useful but is clearly not as relevant. Given a tight training budget, we would prioritize the ``Electronics'' sample over the ``Clothing'' sample. 

It should be noted that simply including samples from categories that are ontologically closest to “Computers” often underperforms in practice, as our offline experiments have shown. This example illustrates the intuition behind domain-aware sampling, yet high-level domain knowledge alone is insufficient to quantify the informativeness of out-of-domain data. This motivates the use of embedding-based and language-model–driven techniques discussed further in Section~\ref{sec:dfm}.
\end{example}

Example \ref{ex:neighbor} highlights the central question addressed in this work: how to select the \emph{best} out-of-domain training samples under a budget constraint in order to optimize EM model performance for each domain. More generally, the structural similarities we intuitively observe between domains--such as between computers and electronics--can be captured by comparing the distributions of their data in a shared high-dimensional space. Distribution-aware methods make it possible to formalize these similarities, guiding the selection of out-of-domain samples that most effectively complement the in-domain data.

Our work makes the following contributions:
\begin{itemize}
\setlength\itemsep{0.5em}
\item \textit{Problem Formulation.} We formalize \emph{Budget-Aware Entity Matching Across Domains} (EMAD), extending EM to account for specialized per-domain models and a training annotation budget that reflects realistic labeling constraints. 

\item \textit{Distribution-aware methods.} We propose novel sampling strategies that leverage embedding-level representations of pairwise EM data to guide out-of-domain selection and improve performance across domains.

\item \textit{The BEACON framework.} We introduce BEACON, a novel budget-aware ensemble framework that employs a dynamic training loop and a dual-PLM architecture, periodically regenerating embeddings and resampling training data to optimize model performance.



\item \textit{Experimental Evaluation.} We adapt EM PLMs (e.g., DITTO~\cite{li2020ditto}), recent EM domain adaptation frameworks (e.g., MFSN~\cite{sun2024mfsn}), recent low-resource EM methods (e.g., Battleship~\cite{genossar2023battleship} and PromptEM\\ ~\cite{promptEM2022}), and recent LLM approaches (e.g., Jellyfish~\cite{zhang2023jellyfish,zhang2024jellyfish}) to establish strong baselines for EMAD, over which we conduct extensive experiments 
showing that BEACON outperforms approaches across different budget constraints.  For this evaluation, we derive nine domain\hyp{}partitioned datasets from WDC~\cite{peeters2023wdc}, as well as two additional datasets derived from the WDC Product Corpus~\cite{primpeli2019wdc} and Abt-Buy~\cite{kopcke2010evaluation}, which we also release to support future research on EMAD. We further provide \textit{Open-source Code and Data.}\footnote{\url{https://github.com/nbpulsone/BEACON}} 

\end{itemize}
The remainder of the paper is organized as follows. Section \ref{sec:related_work} reviews related work and positions our study within the broader context of existing research. Section \ref{sec:prelim} introduces our new problem formulation—\emph{Budgeted Entity Matching Across Domains} (EMAD)--along with key preliminaries. In Section \ref{sec:methods}, we describe methods for solving EMAD, including both basic and distribution-aware approaches. Section \ref{sec:methods} also presents the BEACON framework, which optimizes EM performance under the proposed setting. We present our experimental setup, including the datasets and baselines used in Section ~\ref{sec:experimental_setup}. We report experimental results and analyses in Section \ref{sec:experiments}. Finally, Section \ref{sec:conclusion} concludes the paper and outlines directions for future work. 

\section{Related Work} \label{sec:related_work}
Our work builds on recent advances in learning-based EM, particularly those leveraging deep learning and PLMs. Both the blocking and matching phases of EM have received attention in the deep learning literature. For example, several recent works~\cite{brinkmann2024scblock, javdani2019deepblock, thirumuruganathan2021deepblockds} have explored deep learning approaches for the blocking phase. Since our focus is on the matching phase, we assume that our models operate on the output of any effective blocking method--whether traditional or deep learning-based.

Deep learning for matching was first introduced by DeepER~\cite{ebraheem2017deeper} and later extended to DeepMatcher~\cite{mudgal2018deep}. These works framed EM as a pairwise classification task using deep neural networks and achieved strong performance. More recently, PLM-based matching has become the new standard, with DITTO~\cite{li2020ditto} demonstrating that fine-tuning PLMs such as BERT can significantly improve accuracy. 
Rastaghi et al.~\cite{akbarian2022probing} 
adds robustness-focused optimizations to the DITTO pipeline and Adapterem~\cite{mugeni2023adapterem} enhances computational efficiency. Our work also builds on the DITTO architecture, but operates under a different setting: EM under limited data availability and across multiple domains. While this problem has received comparatively little attention, recent works have explored related challenges in cross-domain and low-resource EM.  


\textbf{Cross-Domain and Domain-Aware EM.}
Recent work has explored domain-aware EM approaches that explicitly account for domain information during model training, though these settings differ from BEACON. For example, Unicorn~\cite{unicorn} proposes a unified mixture-of-experts architecture that supports multiple data-matching tasks by learning shared representations across domains. While Unicorn emphasizes cross-task generalization and zero-shot prediction, BEACON instead focuses on \emph{budget-aware selection of cross-domain training data} for a single EM task. HierGAT~\cite{hiergat} models domain-specific attribute, token, and entity interactions using hierarchical graph attention mechanisms. Whereas HierGAT learns explicit domain-conditioned relational structure, BEACON operates in a PLM embedding space and leverages distribution-aware sampling to allocate a fixed annotation budget. Related survey work (e.g., Paganelli et al.~\cite{paganelliBERTsurvey, paganellimultifacetsurvey}) further analyzes how PLM representations adapt to EM-specific patterns, providing
valuable architectural insights into cross-domain EM that complements BEACON’s domain-aware data selection.

A related line of work focuses on domain adaptation for EM, where labeled data from a source domain is used to improve performance on a different target domain, often with limited or no labeled examples. The idea was first introduced by DAME~\cite{trabelsi2022dame}, which employs a mixture-of-experts approach to capture task-specific information from source domains and transfer it to a separate target domain. DADER~\cite{tu2022dader} similarly explores domain adaptation for EM, but uses feature alignment strategies to make the source domain more informative for the target. Most recently, MFSN~\cite{sun2024mfsn} introduced a framework that explicitly models private and common matching features across domains using multiple encoders, enabling more effective knowledge transfer. We use the enhanced variant, MFSN-FRSE, as a baseline in our work.

While our work shares the high-level goal of improving EM under domain shifts, there are two important distinctions that highlight the novelty of our problem setting. First, we do not assume access to a separate source-domain dataset; instead, we operate within a single large dataset that contains multiple related domains of interest. These domains can be viewed as \emph{micro-domains}--related subdomains (e.g., ``Computers'' and ``Clothing'') within a larger \emph{macro-domain} (e.g., ``e-commerce Products''). In contrast, existing work on cross-domain EM~\cite{trabelsi2022dame, tu2022dader, sun2024mfsn} primarily focuses on transferring knowledge across distinct macro-domains. Second, we assume access to a limited amount of in-domain labeled data and study how to augment it under a constrained sample budget. Rather than transferring representations or model parameters between datasets, our formulation emphasizes \emph{selecting samples across domains within the same dataset}.

\begin{table*}[t]
\centering
\small
\caption{Summary of baseline methods and experimental configurations. \cmark\ and \xmark\ indicate whether a method uses labeled in-domain data, incorporates out-of-domain samples, or requires model training in our experimental setting.}
\label{tab:baseline_summary_table}
\resizebox{\linewidth}{!}{%
\begin{tabular}{lcccc}
\toprule
\revA{\textbf{Method}} &
\revA{\textbf{Uses In-Domain Labels}} &
\revA{\textbf{Uses Out-of-Domain Data}} &
\revA{\textbf{Requires Training}} &
\revA{\textbf{Selection Strategy}} \\
\midrule
\revA{GEN} &
\revA{\cmark} &
\revA{\cmark} &
\revA{\cmark} &
\revA{Random sampling across all domains} \\

\revA{SPEC} &
\revA{\cmark} &
\revA{\xmark} &
\revA{\cmark} &
\revA{Random sampling within target domain} \\

\revA{MFSN~\cite{sun2024mfsn}} &
\revA{\cmark} &
\revA{\cmark} &
\revA{\cmark} &
\revA{Feature-space domain transfer} \\

\revA{Battleship~\cite{genossar2023battleship}} &
\revA{\cmark} &
\revA{\cmark} &
\revA{\cmark} &
\revA{Graph-based uncertainty and centrality} \\

\revA{PromptEM~\cite{promptEM2022}} &
\revA{\cmark} &
\revA{\xmark} &
\revA{\cmark} &
\revA{Prompt tuning with pseudo-label generation} \\

\revA{LLM-based (LLaMA~\cite{dubey2024llama}, Jellyfish~\cite{zhang2023jellyfish})} &
\revA{\xmark} &
\revA{\xmark} &
\revA{\xmark} &
\revA{N/A (Zero-shot inference)} \\
\midrule
\revA{\textbf{BEACON (ours)}} &
\revA{\cmark} &
\revA{\cmark} &
\revA{\cmark} &
\revA{Distribution-aware dynamic resampling} \\
\bottomrule
\end{tabular}

}
\end{table*}

\textbf{Low-Resource and Budget-Constrained EM.} 
Some works have devised methods to perform EM in settings with limited labeled data. PromptEM~\cite{promptEM2022} applies prompt tuning and pseudo-labeling to improve performance for low-resource EM. Other works have applied active learning techniques to EM under limited supervision~\cite{sarawagi2002interactive, bellare2012active, qian2017active, kasai2019low, huang2023active}. The idea was first introduced by Sarawagi and Bhamidipaty~\cite{sarawagi2002interactive}, who demonstrated that a matcher can be effectively trained using an intelligently selected subset of training data. Building on this foundation, Qian et al.~\cite{qian2017active} proposed an active learning framework that learns a diverse set of rules to improve large-scale matching performance. More recent approaches, such as those of Kasai et al.~\cite{kasai2019low} and Huang et al.~\cite{huang2023active}, guide the training of deep learning–based EM models by selecting informative examples. Other works~\cite{genossar2023battleship, de2025gralmatch} explicitly explore fine-tuning PLMs under fixed labeling budgets, but this remains an under-explored area despite its practical relevance. Among these approaches, we adapt a SOTA active learning method for EM, Battleship~\cite{genossar2023battleship}, and low-resource EM method, PromptEM~\cite{promptEM2022}, to the EMAD setting and use them as baselines in our experimental evaluation.

Our work contributes to this line of research by introducing novel sampling strategies designed to maximize performance under strict labeling constraints. However, our approach is not a direct application of active learning or low-resource EM techniques. Instead, we operate in a distinct problem setting that explicitly considers cross-domain structure: given limited supervision in a target domain, we study how to select additional samples to label from \emph{other} domains (and, symmetrically, how out-of-domain supervision can complement scarce in-domain labels) to improve target-domain performance. In contrast, active learning methods typically select the most informative unlabeled samples \emph{within} a single domain for annotation, largely ignoring domain boundaries altogether. 

It is important to recognize the strong connection between cross-domain and low-resource EM--the intersection in which our work resides. In practice, low-resource settings can exploit domain distinctions by selecting samples that best align with existing data distributions, focusing labeling on the most informative samples. Conversely, domain variation often creates low-resource challenges--some domains have ample labeled data while others do not--motivating approaches that transfer knowledge across domains. This natural interplay between limited supervision and domain variation underscores the motivation for our formulation of the EMAD problem, discussed further in Section~\ref{sub:prob_def}. 

\textbf{LLM and Prompt-based EM.}
Recent work~\cite{peeters2023llm, comem} has begun to explore the use of LLMs and prompt-based techniques for EM. A representative example is ComEM~\cite{comem}, which ensembles multiple LLMs and prompting strategies to generate globally consistent EM predictions. These approaches primarily focus on adapting large pretrained models to the EM task, whereas BEACON provides a data selection strategy that can be paired with any trained PLM or LLM-based classifier, including prompt-based systems. In this work, however, we focus on low-resource, budget-aware EM, and leave the integration of BEACON with LLM-based EM approaches for future work.

Table~\ref{tab:baseline_summary_table} situates our methods within the broader EM literature and clarifies how they differ from the baseline approaches in terms of supervision, domain usage, and training requirements.

\section{Preliminaries and Problem Definition} \label {sec:prelim}

We now introduce the notation used 
in the paper (Section \ref{sub:notations_and_blocking}), the standard EM task (Section \ref{sub:em_prelim}), and the extended formulation of Budget-Aware Entity Matching Across Domains (Section \ref{sub:prob_def}). 

\subsection{Notations and Blocking}
\label{sub:notations_and_blocking}
We follow Li et al.~\cite{li2020ditto} and others in defining EM as the task of matching two distinct datasets. Other formulations also exist, where matching is performed within a single dataset ~\cite{sarawagi2002interactive} (i.e., de-duplication). Let $D_{\text{left}}$ and $D_{\text{right}}$ denote the two datasets to be matched, with $|D_{\text{left}}| = l$ and $|D_{\text{right}}| = m$. Each dataset is tabular, consisting of \emph{entities} (or records), where an entity $r$ can be represented as a row of structured information in a table (dataset). In this work, we focus on the \emph{clean-clean} EM setting, which assumes that neither $D_{\text{left}}$ nor $D_{\text{right}}$ contains duplicates internally. In what follows, the set of all possible pairs of entities is given by the Cartesian product $D_{\text{left}} \times D_{\text{right}}$. We refer to these pairs as \emph{candidates}, and the complete set of such pairs as the \emph{candidate set}.

The EM process typically proceeds in two stages: a blocking phase followed by a matching phase. In the blocking phase, a set of lightweight heuristics is applied to eliminate unlikely matches in the \emph{candidate set} and reduce the search space. This yields a subset of candidates $N\subset D_{\text{left}} \times D_{\text{right}}$ such that $|N| \approx \max(\mathcal{O}(l), \mathcal{O}(m))$, which avoids the $\mathcal{O}(l\times m)$ complexity of exhaustive pairwise comparison. While blocking is essential to scalability, this paper focuses on improving the downstream matching phase.

We assume that the set $N$ can be partitioned into $j$ disjoint subsets--each corresponding to a different domain--such that:
\[
N = n_1 \cup n_2 \cup \dots \cup n_j, \quad n_i \cap n_k = \emptyset \text{ for } i \ne k
\] 

In this work, we focus on product data, where the domain corresponds to product categories (e.g., "Computers", "Clothing", etc.) or brands (e.g., "Apple", "Samsung", etc.). 

\subsection{Entity Matching} \label{sub:em_def}
\label{sub:em_prelim}
Given a filtered candidate set $N$, the goal of the matching phase is to determine which candidates contain matching entities. This is formulated as a binary classification task: for each candidate $(r_1, r_2)_i \in N$, we assign a label $y_i \in \{0, 1\}$, where $y_i = 1$ if $r_1$ and $r_2$ are matching entities, and $0$ otherwise. We refer to a candidate together with its corresponding label as a \emph{sample}.

We adopt PLMs such as BERT~\cite{devlin2019bert} (or RoBERTa~\cite{liu2019roberta} and DistilBERT~\cite{sanh2019distilbert}) to perform the classification. A PLM refers to a transformer\hyp{}based model trained on a large corpus of text, which can be fine-tuned for downstream tasks such as EM. Following the DITTO framework~\cite{li2020ditto}, we serialize an entity $r$ with $p$ attributes as:
\[
\texttt{[COL]} \; \texttt{attr}_1 \; \texttt{[VAL]} \; \texttt{val}_1 \cdots \; \texttt{[COL]} \; \texttt{attr}_p \; \texttt{[VAL]} \; \texttt{val}_p
\] 
and each candidate as:
\[
\texttt{[CLS]} \; r_1 \; \texttt{[SEP]} \; r_2 \;
\] 
This serialized string is passed through the PLM. The \texttt{[SEP]} and \texttt{[CLS]} tokens are special tokens included in the input. The \texttt{[SEP]} token indicates the boundary between the left and right entities in the serialized string. The output embedding corresponding to the \texttt{[CLS]} token is treated as a summary vector for the candidate. Importantly, the label indicating whether a pair is a match or non-match is not included in the serialized string used to form the embedding representation. This vector is then passed through a fully connected layer, followed by a softmax or sigmoid activation, to produce the match prediction $\hat{y}_i$.

\subsection{Problem Definition}
\label{sub:prob_def}
We now extend the formulation to the \emph{Budget-Aware Entity Matching Across Domains} (EMAD) setting. We define the set of domains to be $\mathcal{D}$, where $|\mathcal{D}| = j$. Instead of training a single matcher on $N$ for all domains, we train $j$ domain-specific matchers using the sets $X_1, X_2, \dots, X_j$, where for $i \in \mathcal{D}$, each $X_i$ is a training set for domain $i$. The na\"ive way to address this is use only domain-specific data for each training set, where $X_i = n_i, \forall i$ (see Section ~\ref{sub:spec}). 

While it may also be attractive to train a single matcher for all domains (i.e., $X_i = N, \forall i$), there are strong motivations for using domain-specific matchers. We illustrate these motivations with the following example:

\begin{example} \label{ex:ds_motivation}
Consider the task of matching entities across two product domains: ``Shoes'' and ``Jewelry.''
For shoes, EM often depends on a small set of attributes with highly standardized values.
For instance, two Nike Air Zoom Pegasus 40 running shoes (US size 10, Style Code \texttt{DV3853-401}, retail price \$130) can be matched with high confidence if their \emph{Brand}, \emph{Size}, and \emph{Style Code} attributes align across both records.\footnote{Please find the full item description~\cite{AmazonNikeZoom}}

Conversely, matching jewelry items such as gold rings requires reasoning over more descriptive and less standardized attributes--e.g., \emph{material composition}, \emph{metal}, \emph{gemstone type}, and \emph{carat weight}--where subtle differences in any one of these attributes may alone determine whether two rings refer to distinct products.\footnote{Please find the full item description~\cite{AmazonRingSolitaire}}

Moreover, the interdependence between attributes differs across domains: in footwear, the \emph{Style Code} is meaningful only in conjunction with the \emph{Brand}, whereas in jewelry, \emph{Brand} and \emph{Size} alone do not provide much matching insight without additional context.
\end{example}

Example~\ref{ex:ds_motivation} highlights that a single matcher is unlikely to capture the diverse decision patterns observed across domains--a phenomenon also noted in other machine learning tasks such as recommendation systems~\cite{jiang2022adaptivedomain, simioni2024automated}. By contrast, a matcher trained on a specific domain can effectively capture domain-specific nuances, even when out-of-domain candidates are included in the training set. Prior work in domain adaptation~\cite{tu2022dader} has demonstrated that models trained on domain-relevant data achieve improved performance within their target domain. Unlike these approaches, however, our formulation builds on this insight by explicitly studying how to \emph{borrow out-of-domain signal under a fixed annotation budget} to complement limited in-domain data, with the goal of training specialized domain-specific models rather than assuming that a single unified model will suffice.


We assume that each domain is defined such that both entities in any candidate belong to the same domain (e.g., both records are from the "Computers" category). For each domain $i$, we are given a \emph{training budget} $\beta$ that constrains the total number of candidates in $X_i$. These candidates may be drawn from both in-domain data ($n_i$) and out-of-domain data ($N \setminus n_i$).

\begin{definition}[Budget-Aware Entity Matching Across Domains (EMAD)]
Given a domain partition of $N = n_1 \cup n_2 \cup \dots \cup n_j$ over the domains $\mathcal{D}$ of the candidate set and a training budget $\beta$, the goal of EMAD is, for each domain $i \in \mathcal{D}$, to construct a budgeted training set
\[
X_i = n_i^{*} \cup S_i^{\text{out}}, \quad \text{where } |X_i| = \beta, \quad S_i^{\text{out}} \subseteq (N \setminus n_i),
\]
such that the resulting model achieves high matching performance on domain $i$.
\end{definition}

Here, $n_i^{*}$ denotes a subsampled or oversampled version of $n_i$, and $S_i^{\text{out}}$ is a set of out-of-domain samples selected according to a given sampling strategy. This formulation allows us to investigate how different sampling techniques affect domain-specific model performance under realistic budget constraints.


\section{Budget-Aware Entity Matching Across Domains} \label{sec:methods}

We now introduce a suite of methods to address the EMAD 
problem, along with complementary techniques to improve PLM-based EM under budget constraints. 

Section~\ref{sec:basic_methods} presents foundational baseline models, each motivated by simple but effective heuristics for sample selection. Section~\ref{sec:dfm} expands on this by introducing models that perform distribution\hyp{}aware sample selection, using embedding-based techniques to align out-of-domain data with in-domain distributions. Finally, Section~\ref{sec:beacon} introduces BEACON--our central contribution--an ensemble-based framework that incorporates a dynamic resampling procedure to maximize EM performance across domains.


\subsection{Basic Models} \label{sec:basic_methods}
We begin by introducing basic models for EMAD. 
Although simple, these models provide strong baselines 
as they are grounded in natural intuitions that also serve as the foundation for our more advanced models. Throughout this section and Section~\ref{sec:dfm}, we employ a common operator, $\texttt{RandomSelect}(A, b)$, which randomly selects $b$ items from a set $A$, where $b$ is an integer such that $b \leq |A|$. This operator defines the random sampling mechanism that is used in several of our formulations.

\subsubsection{The General Model (GEN)}

The General Model (GEN) generates $X_i$ by randomly sampling from the candidate set, without regard to domain. For example, if we are training a model to match WDC ``Computers'' pairs, we sample across all available domains (e.g., ``Electronics,'' ``Clothing,'' etc.) to fill the budget, including candidates from the target domain itself (see Example ~\ref{ex:em_example}). GEN corresponds to the idea of training a single matcher across all domains, as discussed in Section~\ref{sub:prob_def}. By randomly sampling across all domains, GEN simulates a unified model trained without regard to domain boundaries, while still respecting the budget constraint. GEN is motivated by the hypothesis that broad domain coverage may expose the model to a more diverse range of attribute patterns, potentially improving its generalization to unseen data. In practice, GEN represents the most domain-agnostic baseline.

Formally, for $1 \leq i \leq j,$ we have
\begin{equation} \label{eq:gen}
X_{i} = \texttt{RandomSelect}(N, \beta)
\end{equation}

Given a budget $\beta$, GEN draws $\beta$ samples from the candidate set at random to construct a training set for each domain. Thus, the training set for each domain becomes a random mixture of in-domain ($n_i$) and out-of-domain ($S_{i}^{\text{out}}$) candidate pairs.

\subsubsection{The Domain-Specific Model (SPEC)} \label{sub:spec}

The Domain-Specific Model (SPEC) generates $X_i$ exclusively using samples from the target domain. Given the differences in the EM process across domains, we consider the effect of training solely on domain-specific data--forcing the model to focus on the unique characteristics of matching within a given domain. Not only do domains differ in their matching logic and attribute structure, but they can also vary in the availability of training data. As shown in Table~\ref{tab:wdc_dataset_stats}, certain domains can contain orders of magnitude more samples than others.

\begin{figure}[t]
  \centering
  \includegraphics[width=0.7\textwidth]{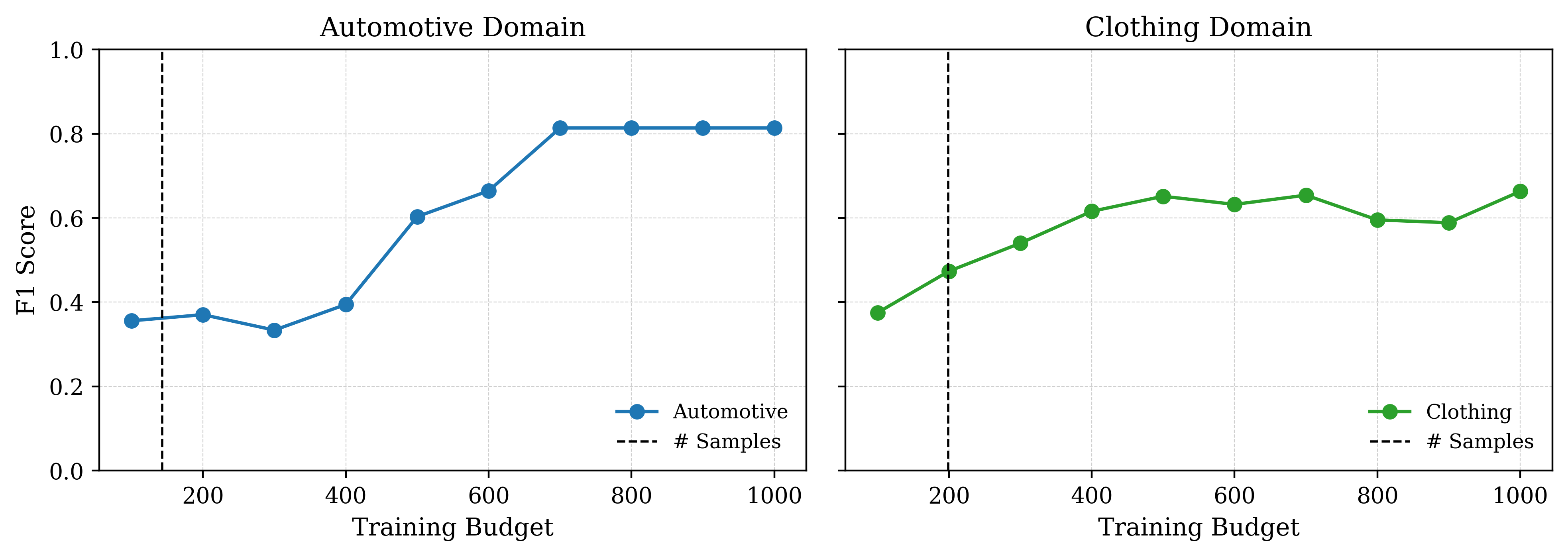}
  \caption{\textbf{Effect of Oversampling on PLM-based Entity Matching.}
  We fine-tune RoBERTa on two small product domains--\emph{Automotive} and \emph{Clothing}--from the WDC dataset using progressively larger training budgets that permit oversampling.}
  \label{fig:oversampling_demo}
\end{figure}

For SPEC, if a domain contains fewer than $\beta$ samples, we oversample (with replacement) from the available data. As depicted in Figure~\ref{fig:oversampling_demo}, oversampling--especially for smaller domains--leads to steady performance improvements that tend to plateau. Conversely, with smaller budgets, the domain may contain more than $\beta$ samples. While state-of-the-art active learning methods~\cite{kasai2019low, huang2023active, genossar2023battleship} have shown that fine-tuning on a selected subset of the available training data can improve matching performance, for SPEC and the models that follow, we instead opt to randomly down-sample to keep the focus of this paper on out-of-domain sampling.

Formally, for $1 \leq i \leq j$, we define:
\begin{equation} \label{eq:spec}
X_i =
\begin{cases}
n_i^{\lfloor \frac{\beta}{|n_i|} \rfloor} \cup \texttt{RandomSelect}(n_i, \beta \bmod |n_i|), & \text{if } |n_i| \leq \beta, \\
\texttt{RandomSelect}(n_i, \beta), & \text{otherwise.}
\end{cases}
\end{equation}

Given a budget $\beta$, SPEC randomly draws $\beta$ samples from $n_i$ (with replacement if necessary) to construct a training set composed entirely of samples from the relevant domain. Intuitively, Equation (\ref{eq:spec}) formalizes the idea that the SPEC model is restricted to in-domain information, serving as the baseline for performance when no cross-domain data is leveraged. 

\subsection{Distribution-Aware Models} \label{sec:dfm}
We now introduce models that approach EMAD using distribution-aware techniques.
Modern language models enable serialized candidate pairs to be represented as high-dimensional vectors, commonly referred to as \emph{embeddings}~\cite{zeakis2025depth} (see Section~\ref{sub:em_def}).
These embeddings capture semantic information about the textual content of the records; however, they do not encode label information, ensuring that the selection process is driven by unsupervised characteristics of the data. The closer two embeddings are in this space, the more semantically similar their corresponding candidate pairs. 

Because domains differ in vocabulary, attribute composition, and matching logic, their embeddings often form clustered regions. While these regions may overlap--particularly for semantically related domains--the overall structure tends to reflect meaningful domain-level distinctions. We refer to the shape and spread of these clustered regions as the \emph{distribution} of a domain’s embeddings--a mathematical characterization of how its candidate pairs are arranged, reflecting attribute correlations, dependencies, and domain-specific terminology. By leveraging these distributions, we can compare domains quantitatively and reason about their similarity. This, in turn, enables \emph{distribution-aware} sampling strategies that selectively incorporate out-of-domain samples whose embeddings align with the in-domain distribution, forming the basis of the novel sampling methods we introduce in this work.


Given a domain $i \in \mathcal{D}$, the goal is to select out-of-domain samples from $N \setminus n_i$ so as to align the distribution of the in-domain samples $n_i$ with a target distribution. Because the distributions vary across domains, the appropriate target distribution may differ from one domain to another. The following models make different assumptions about the target distribution, which guide the selection of out-of-domain samples. 

\subsubsection{The Nearest Neighbors Model (NN)}

The Nearest Neighbors Model (NN) first includes all in-domain samples from $n_i$ in $X_i$, similar to SPEC. In the majority of cases where the budget allows for more samples (e.g., $n_i<\beta$), the model then selects additional samples from other domains whose embeddings are closest to the domain-specific centroid under a distance metric (e.g. cosine distance). Intuitively, these samples are those whose inclusion alters the domain-specific centroid the least. NN assumes that the domain-specific embeddings already provide a good representation of the domain’s distribution.

Formally, for $1 \leq i \leq j$, we define:
\begin{equation}
X_i =
\begin{cases}
\texttt{RandomSelect}(n_i, \beta), & \text{if } |n_i| \geq \beta, \\
n_i \cup \text{NN}_{\beta - |n_i|}(N \setminus n_i, \mu_i), & \text{otherwise},
\end{cases}
\end{equation}
where $\mu_i$ is the centroid embedding of $n_i$, and $\text{NN}_k(A, \mu_i)$ denotes the set of $k$ samples in $A$ whose embeddings are nearest to $\mu_i$.

Given a set of embeddings $A$, we define the NN selector as:
\begin{equation}
\text{NN}_k(A, \mu_i) =
\operatorname*{Top\text{-}k}_{x \in A}
\big(
\cos(\mu_i, x)
\big),
\end{equation}
which represents the $k$ candidate pairs from $A$ with the highest cosine similarity to $\mu_i$, i.e., the $k$ elements $x \in A$ that maximize $\cos(\mu_i, x)$ in descending order. The $\operatorname{Top\text{-}k}$ operator returns the $k$ highest-scoring elements, where ties, if any, are broken arbitrarily.
\subsubsection{The Train-Validation Distribution Fitting Model (TVDF)}

\begin{figure}
  \centering
  \includegraphics[width=0.75\textwidth]{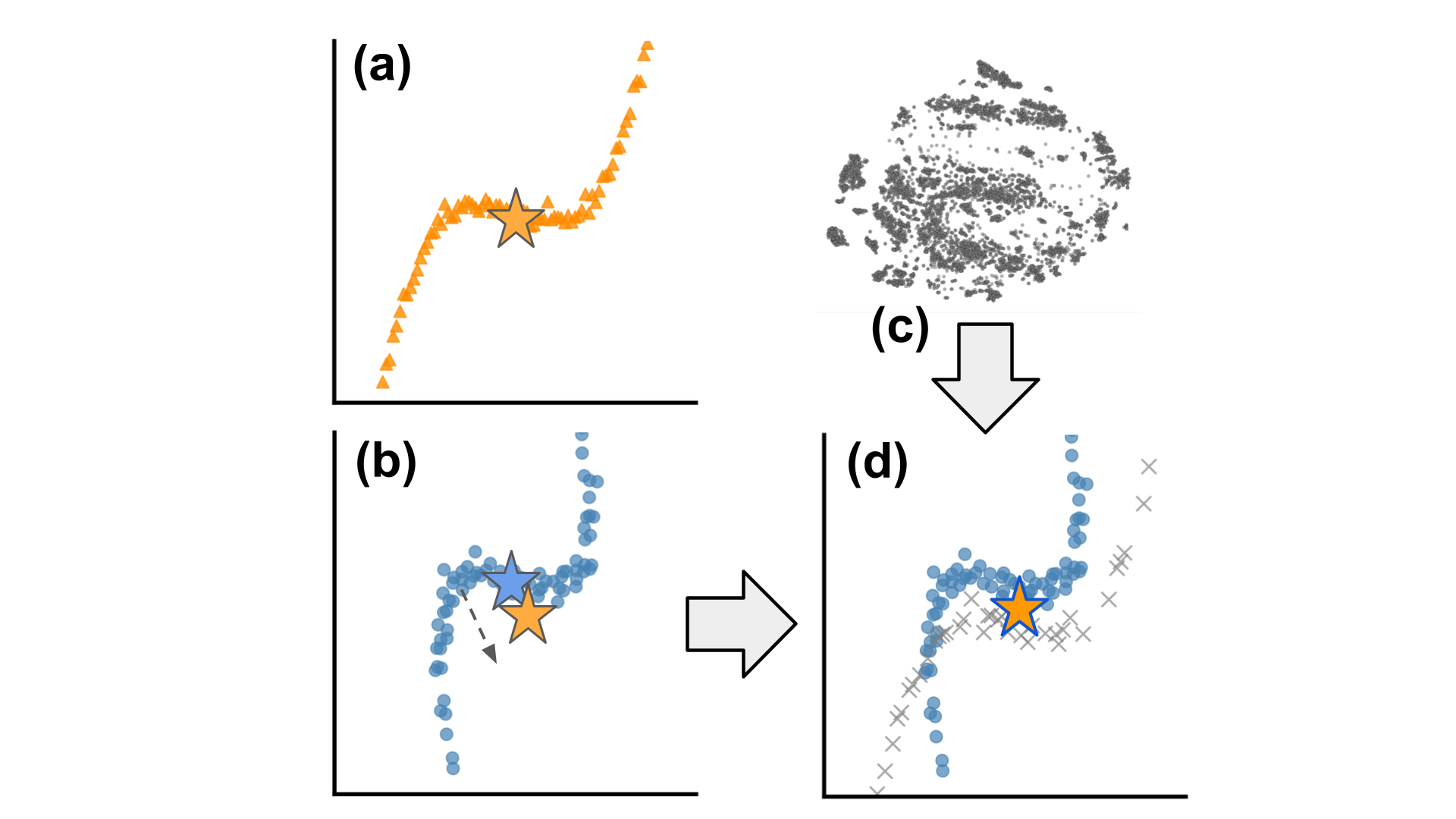}
  \caption{\textbf{Illustration of the Train–Validation Distribution Fitting (TVDF) sampling procedure.} (a) Validation set embeddings with centroid $\mu_{\text{val}}$ (orange star). (b) In-domain training embeddings for domain $i \in \mathcal{D}$ with centroid $\mu_i$ (blue star) and overlaid $\mu_{\text{val}}$. (c) Out-of-domain samples are ranked by the TVDF selector based on their contribution to aligning $\mu_i$ with $\mu_{\text{val}}$. (d) Selected samples are added to the training set, resulting in closer alignment between the training and validation centroids.}
  \label{fig:tvdf_example}
\end{figure}

The Train-Validation Distribution Fitting Model (TVDF) shares a similar setup with NN. It first includes all available domain-specific samples from the target domain in $X_i$. When $n_i<\beta$, it selects additional samples from other domains whose embeddings bring the training distribution closer to that of the target domain's validation set.

More concretely, we compute the centroid of the target domain's training embeddings and the centroid of its validation embeddings. Additional samples are drawn from other domains such that, when added to the training set, they minimize the distance between the updated training centroid and the validation centroid. In this view, TVDF assumes the current in-domain labeled data are insufficient to represent the target domain’s embedding distribution, and therefore leverages the validation set as a proxy for the (unseen) test distribution.

Formally, for each domain $i \in \mathcal{D}$, we define the centroids of the training and validation embeddings as:
\begin{equation}
\mu_i = \frac{1}{|n_i|}\sum_{x \in n_i} x,
\qquad
\mu_{\text{val}_i} = \frac{1}{|v_i|}\sum_{x \in v_i} x,
\end{equation}
where $v_i$ denotes the validation embeddings for domain $i$. 

Given a total training budget $\beta$, the budgeted training set $X_i$ is constructed as:
\begin{equation}
X_i =
\begin{cases}
\texttt{RandomSelect($n_i, \beta$)}, & \text{if } |n_i| \geq \beta, \\[4pt]
n_i \cup \text{TVDF}_{\beta - |n_i|}(N \setminus n_i, \mu_i, \mu_{\text{val}_i}), & \text{otherwise.}
\end{cases}
\end{equation}

For each candidate pair $x \in N \setminus n_i$, let the updated centroid after adding $x$ be:
\begin{equation}
\mu_{n_i \cup x} = \frac{|n_i| \cdot\mu_i + x}{|n_i| + 1}.
\end{equation}
Given a set of embeddings $A$, we then define the TVDF selector as:
\begin{equation}
\text{TVDF}_k(A, \mu_i, \mu_{\text{val}_i}) =
\operatorname*{Top\text{-}k}_{x \in A}
\big[
\cos(\mu_{n_i \cup x}, \mu_{\text{val}_i}) -
\cos(\mu_i, \mu_{\text{val}_i})
\big],
\end{equation}
which returns the $k$ candidate pairs from $A$ that most increase the cosine similarity between the updated training centroid and the validation centroid. Figure~\ref{fig:tvdf_example} illustrates the TVDF selector in a toy example; for simplicity, candidates are shown in a two-dimensional embedding space with the training and validation centroids overlaid.

\subsubsection{K-Center Greedy Sampling (KCG)}

\begin{figure}
  \centering
  \includegraphics[width=0.7\textwidth]{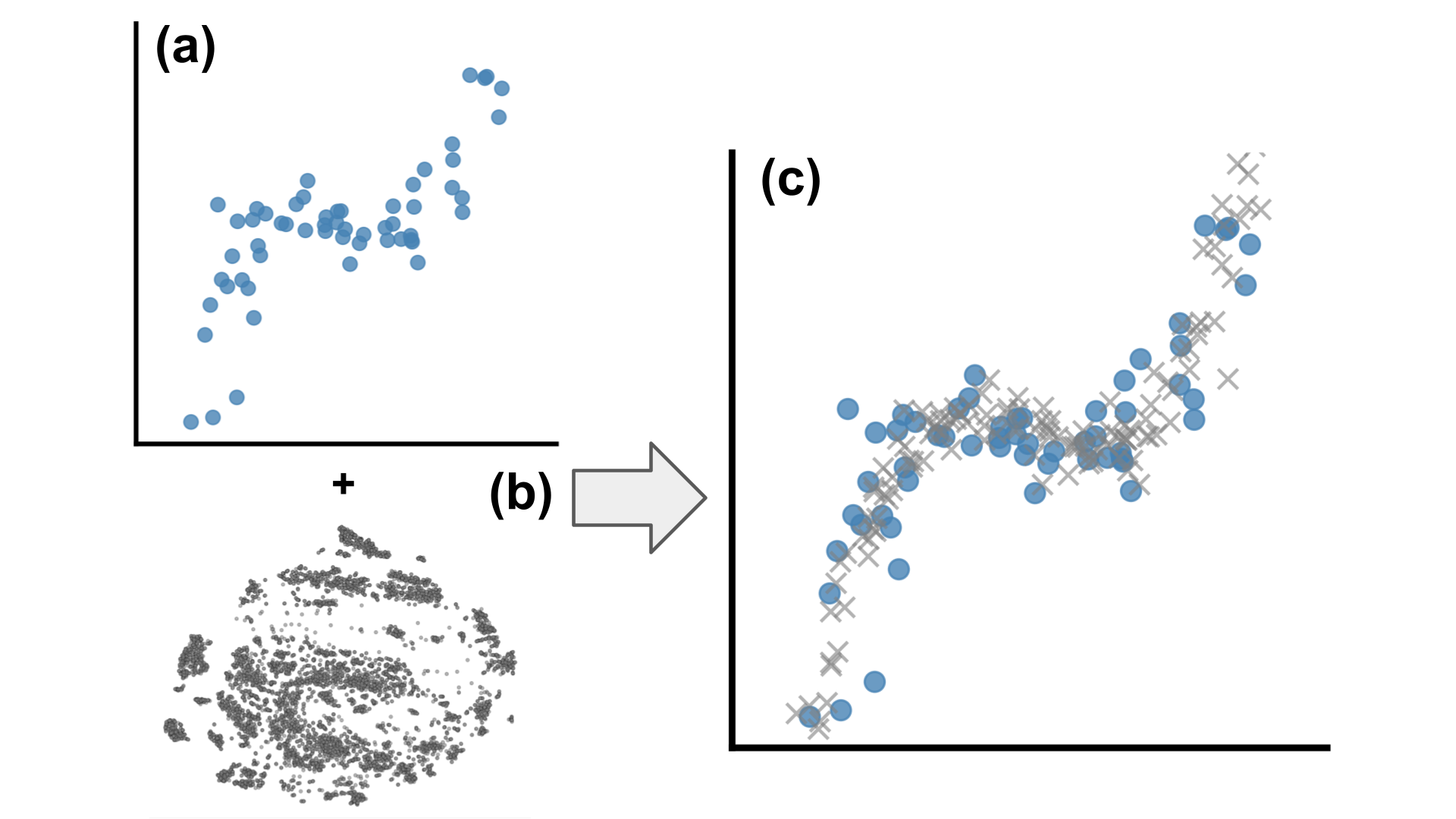}
  \caption{\textbf{Illustration of the K-Center Greedy (KCG) sampling procedure.} (a) In-domain embeddings $n_i$ serve as the initial centers. (b) Out-of-domain samples are ranked by the KCG selector based on their distance to the nearest in-domain center. (c) Selected samples are added to the training set to fill gaps in the embedding space, increasing overall coverage.}

  \label{fig:kcg_example}
\end{figure}

The K-Center Greedy (KCG) model treats the selection of out-of-domain samples as a $k$-center optimization problem. 
Here, the embeddings of the in-domain samples $n_i$ serve as the initial centers, and the goal is to select samples from $N \setminus n_i$ 
such that the maximum distance from any sample to its nearest center is minimized. 
This corresponds to a farthest-first traversal strategy that incrementally adds the most distant points to maximize coverage in the embedding space.

KCG assumes that the in-domain data already provides a strong local representation of the domain. 
Its objective is to improve generalization by incorporating diverse out-of-domain samples that fill gaps in this representation.

Formally, for each domain $i \in \mathcal{D}$, we define the training set as:
\begin{equation}
X_i =
\begin{cases}
\texttt{RandomSelect($n_i, \beta$)}, & \text{if } |n_i| \geq \beta, \\[4pt]
n_i \cup \text{KCG}_{\beta - |n_i|}(n_i, N \setminus n_i), & \text{otherwise.}
\end{cases}
\end{equation}

For each candidate $x \in N \setminus n_i$, its distance to the nearest in-domain center $c$ is defined as:
\begin{equation}
d(x, n_i) = \min_{c \in n_i} \big( 1 - \cos(x, c) \big).
\end{equation}
The KCG selector then chooses the $k = \beta - |n_i|$ candidates that maximize this distance:
\begin{equation}
\text{KCG}_k(n_i, A) =
\operatorname*{Top\text{-}k}_{x \in A} \; d(x, n_i),
\end{equation}
This is solved greedily by repeatedly adding the farthest embedding in $A$ and updating the distance to the nearest center, yielding a 2-approximation to the optimal $k$-center solution under cosine distance. Figure~\ref{fig:kcg_example} illustrates the KCG selection process in a simplified 2-dimensional embedding space.


\subsection{BEACON} \label{sec:beacon}

\begin{figure}[htbp]
  \centering
  \includegraphics[width=\linewidth]{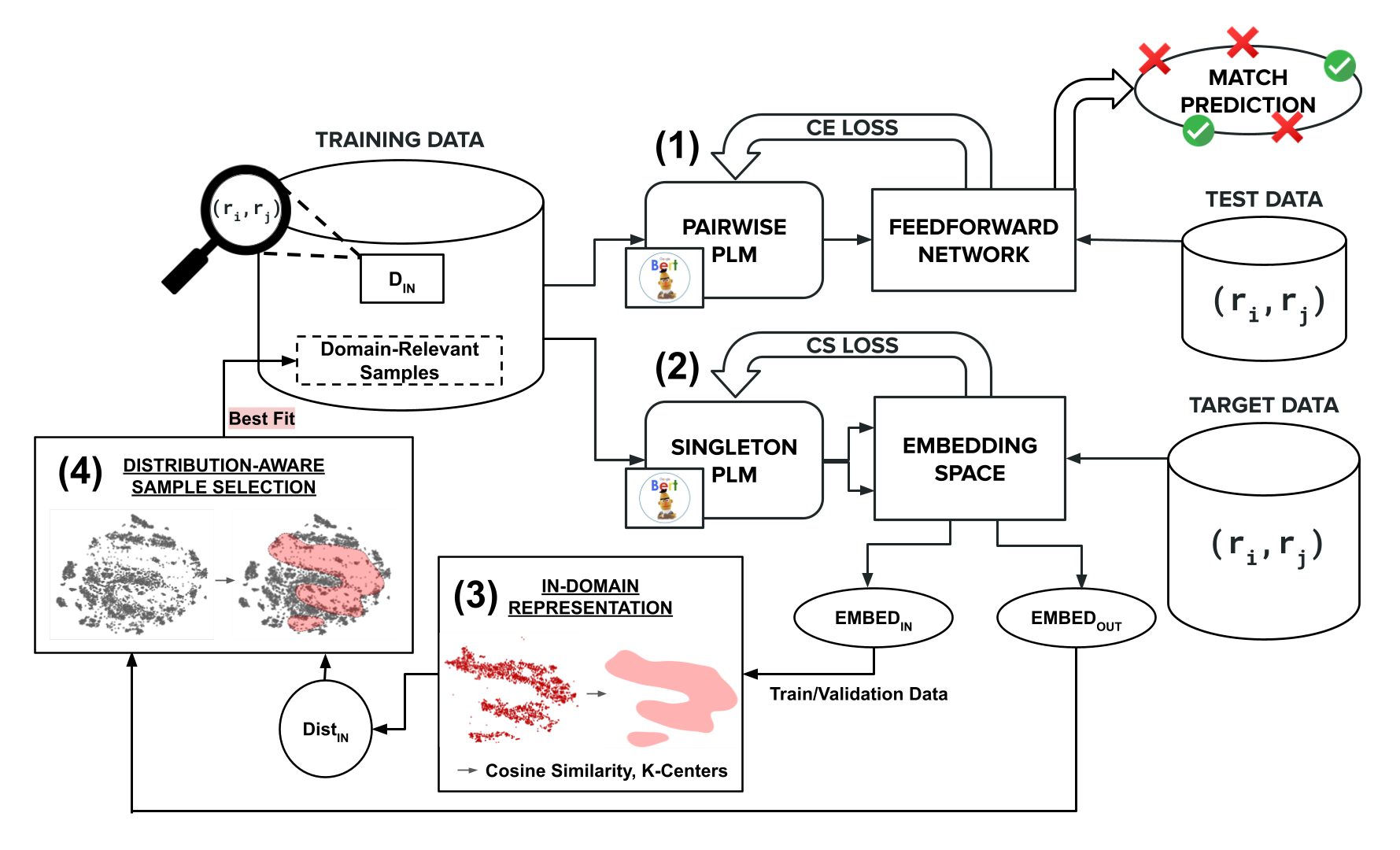}
  \caption{\textbf{An Illustration of the BEACON framework.} (1) A pairwise PLM is fine-tuned on budgeted in- and out-of-domain candidates for EM. (2) A singleton PLM learns embedding representations ($EMBED_{IN}, EMBED_{OUT}$) to guide sample selection. (3) In-domain embeddings form a distribution representation ($Dist_{IN}$) based on cosine similarity or $k$-centers. (4) Distribution-aware sampling selects out-of-domain examples to update the training set before fine-tuning resumes.}
  \label{fig:beacon_pipeline}
\end{figure}

We present \textbf{BEACON} (\textbf{B}udget-aware \textbf{E}ntity matching \textbf{AC}r\textbf{O}ss domai\textbf{N}s), our proposed framework for optimizing EMAD under budget constraints. 
Figure~\ref{fig:beacon_pipeline} presents an overview of the BEACON framework, highlighting its three major components: 
(1) an \textit{Ensemble of Models} (Section~\ref{sec:ensemble}), 
(2-4) a \textit{Dynamic Training Loop} (Section~\ref{sec:dynamic_loop}), and 
(3–4) a \textit{Resampling Mechanism} (Section~\ref{sec:resampling}). 
Notably, BEACON’s dynamic training loop and distribution-aware resampling mechanism constitute novel components of the framework which have not been explored in prior cross-domain EM research. Together, these components enable iterative refinement of the training data and progressively improve EM performance across domains.

\subsubsection{The Ensemble of Models} \label{sec:ensemble}

Following prior work on ensemble-based EM systems~\cite{ding2024setemensemble, low2024emensemble}, we propose an ensemble model to improve classification performance under the EMAD setting. Ensemble methods are widely recognized for enhancing predictive robustness, with even simple aggregation strategies such as soft voting proving effective across diverse classification tasks~\cite{dietterich2000ensemble, kuncheva2014ensemble}.

Once several models are trained, their predictions can be aggregated to form a more accurate consensus. Given $k$ trained models, let $q \in [1, k]$ index a model. We assume access to a validation set $v_i$ for each domain $i \in \mathcal{D}$ and measure the validation F1 performance of each model as $F1_{\text{VAL}, q}$. We adopt a \textit{weighted soft voting} scheme, in which the confidence outputs $Conf_q$ from each model are combined using weights proportional to their validation F1 scores. This provides a principled way to emphasize models that perform best during validation. Formally, the final ensemble confidence $Conf_f$ for a given candidate pair is computed as:

\begin{equation} \label{eq:probf}
Conf_f = \frac{\sum_{q=1}^{k} Conf_q \cdot F1_{\text{VAL}, q}}{\sum_{q=1}^{k}F1_{\text{VAL}, q}}
\end{equation}

\subsubsection{The Dynamic Training Loop} \label{sec:dynamic_loop}

As mentioned in Section~\ref{sec:introduction}, our models build on the state-of-the-art EM framework DITTO~\cite{li2020ditto}, which fine-tunes a PLM for EM tasks. Unlike prior frameworks~\cite{mudgal2018deep, li2020ditto, sun2024mfsn} that train on a static dataset, BEACON introduces a \textit{dynamic training loop} that adaptively resamples data using updated embeddings. This mechanism improves embedding quality throughout training--specifically for the NN, TVDF, and KCG models, which rely on meaningful structure in the embedding space. The cycle formed by steps (2-4) in Figure~\ref{fig:beacon_pipeline} illustrates the training loop.

While PLMs can produce reasonable embeddings out of the box, fine-tuning them on the EM task yields more accurate, domain-specific representations. Accordingly, BEACON pauses training at intermediate checkpoints to regenerate embeddings using the current model state, and then resamples the training data based on the updated embeddings. This enables better alignment between the training distribution and the embedding geometry.

To support this dynamic resampling process, BEACON integrates two coordinated PLMs with complementary objectives:
\begin{itemize}
    \item A \emph{pairwise} PLM, fine-tuned on labeled candidate pairs using cross-entropy loss for the core EM task.
    \item A \emph{singleton} PLM, trained on individual entities using cosine similarity loss, whose role is to generate improved embeddings for the resampling procedure.
\end{itemize}

We also experimented with a simplified setup that used the singleton PLM for both training and embedding generation, but this consistently underperformed relative to the dual-PLM design. The separation allows the pairwise PLM to specialize in the matching task, while the singleton PLM provides embeddings that better capture entity-level similarity to guide resampling. This dual-PLM architecture is novel in the EM literature, explicitly decoupling the supervision and representation roles of PLMs.

\subsubsection{Resampling Techniques} \label{sec:resampling}

We now focus on the resampling stage of the dynamic training loop, which is crucial for enabling our models to maintain a pool of increasingly relevant training samples. At this point in the loop, we assume the model has paused at a checkpoint from which updated embeddings can be extracted for both in-domain and out-of-domain samples. The three methods that utilize this resampling process--NN, TVDF, and KCG--share a common structure and goal: to dynamically select out-of-domain examples that most effectively complement the in-domain data.

Figure~\ref{fig:beacon_pipeline} (steps 3-4) illustrates the shared workflow. Each method first constructs a representation of the in-domain distribution (e.g., a centroid or a set of embedding centers). Out-of-domain samples are then ranked by how well they align with this representation, using the scoring criteria specific to each method. Finally, the top-$k$ samples are selected to be merged with the in-domain training set, and the training loop resumes. Repeating this process periodically allows the model to gradually refine its understanding of the embedding space and sample distribution during training.

\section{Experimental Setup} \label{sec:experimental_setup}

We conduct a comprehensive set of experiments to evaluate our proposed methods against several baselines, including LLM-based approaches, under the EMAD setting. In Section ~\ref{sub:datasets}, we discuss the datasets we use for our experiments. We follow with implementation details in Section ~\ref{sub:implementation_details}, and introduce the baselines we use for our experiments in Section ~\ref{sub:baselines}.

\subsection{Datasets} \label{sub:datasets}

To evaluate EMAD, we derive datasets from three widely used EM benchmarks: the WDC Multi-Dimensional Benchmark~\cite{peeters2023wdc}, the WDC Products Corpus~\cite{primpeli2019wdc}, and Abt-Buy~\cite{kopcke2010evaluation}.

\textbf{Domain Partitioning.} Following Example~\ref{ex:neighbor}, we partition each benchmark into multiple domain-specific EM datasets based on a semantic domain attribute (e.g., \emph{product category} or \emph{brand}). We retain only domains containing at least 100 labeled samples to ensure sufficient data for domain-specific fine-tuning and evaluation. When partitioning by domain, we preserve the original train, validation, and test splits provided by each benchmark rather than artificially rebalancing them. These splits reflect the inherent heterogeneity of real-world e-commerce data, including variation in sample availability and class balance across domains. The distribution-aware methods introduced in Section~\ref{sec:dfm} aim to mitigate these disparities by adaptively sampling informative out-of-domain samples that can compensate for data scarcity or skewed class ratios.

\textbf{The WDC Multi-Dimensional Benchmark.} WDC is a large collection of product-based datasets that vary along three controlled dimensions: (1)~the percentage of corner cases (i.e., difficult-to-match samples), (2)~the percentage of unseen entities in the test set, and (3)~the size of the development set. Unlike 
other EM benchmarks, WDC assigns a \textit{gold-standard product category label} to each record, enabling a definitive partition of the data into domains. This property makes WDC uniquely suited for evaluating domain-aware EM frameworks such as BEACON.

We use the \textit{large} versions of the WDC datasets to guarantee a sufficiently large sample pool for studying different budget constraints. Varying the remaining two dimensions of WDC (corner case percentage and unseen-entity ratio) yields a total of \textbf{nine datasets} to use throughout our experiments. In our main experiments, we fix the corner case percentage at 50\% and vary the unseen-entity percentage in the test set (0\%, 50\%, and 100\%). Partitioning these datasets by \emph{product category} yields between 11 and 13 domain-specific subsets, depending on the variant.

\begin{table}[t]
\centering
\caption{Average WDC Dataset Statistics Across Domains. Each value is averaged over all domain partitions.}
\label{tab:wdc_dataset_stats}
\resizebox{0.6\textwidth}{!}{%

\begin{tabular}{l|cc|cc|cc}
\toprule
Split & \multicolumn{2}{c}{Train} & \multicolumn{2}{c}{Validation} & \multicolumn{2}{c}{Test} \\
Statistic & Samples & \% Pos & Samples & \% Pos & Samples & \% Pos \\
Domain &  &  &  &  &  &  \\
\midrule
Automotive & 120 & 57.4 & 32 & 23.7 & 23 & 26.0 \\
Cameras & 1317 & 57.7 & 204 & 17.7 & 172 & 18.8 \\
Cell Phones & 151 & 69.5 & 24 & 36.8 & 48 & 24.5 \\
Clothing & 219 & 58.2 & 28 & 21.2 & 37 & 23.3 \\
Computers & 6473 & 51.2 & 1174 & 14.6 & 790 & 15.6 \\
Home/Garden & 60 & 47.5 & 25 & 23.9 & 125 & 20.9 \\
Jewelry & 526 & 42.4 & 183 & 17.8 & 318 & 17.0 \\
Movies & 58 & 51.7 & 15 & 26.7 & 48 & 18.9 \\
Instruments & 266 & 53.4 & 54 & 20.4 & 100 & 19.1 \\
Office & 1800 & 47.2 & 467 & 16.4 & 496 & 16.5 \\
Electronics & 2759 & 53.5 & 483 & 17.0 & 391 & 18.0 \\
Sports & 1715 & 59.4 & 256 & 18.1 & 211 & 19.9 \\
Tools & 129 & 77.9 & 14 & 41.4 & 29 & 28.7 \\
Toys & 139 & 48.9 & 27 & 18.5 & 25 & 20.0 \\
\midrule
Total & 15497 & 52.8 & 2941 & 16.6 & 2724 & 17.6 \\
\bottomrule
\end{tabular}
}
\end{table}

Table~\ref{tab:wdc_dataset_stats} reports average statistics across the nine WDC variants after domain partitioning. The results show substantial variation in both sample counts and class balance across domains. Some categories, such as \textit{Computers}, contain thousands of labeled examples, while others have only a few hundred. Likewise, the proportion of positive pairs varies widely--ranging from less than 50\% in domains like \textit{Jewelry} to over 70\% in \textit{Tools}. These discrepancies introduce significant distribution shifts in both sample availability and label composition, which directly affect the difficulty of training domain-specific models. 

\textbf{The WDC Product Corpus}. To avoid overfitting our conclusions to datasets with relatively small annotation pools, we additionally include the \emph{medium} variant of the WDC Products Corpus~\cite{primpeli2019wdc} in our main experimental results. This dataset consists of four product domains (``Cameras'', ``Computers'', ``Shoes'', and ``Watches''), each with substantially larger annotation pools (approximately 4k--6k labeled pairs). These larger domains complement the other WDC variants by enabling an evaluation of EMAD in settings where in-domain supervision is already sufficient for fine-tuning, and where the role of out-of domain selection is more nuanced.

\textbf{The Abt-Buy Benchmark.} The Abt-Buy benchmark~\cite{kopcke2010evaluation} is a widely used EM dataset that aligns product listings from Abt.com and Buy.com. While it shares the e-commerce setting of WDC, Abt-Buy differs in that its attributes are predominantly textual and less uniformly structured. This characteristic allows us to evaluate BEACON in scenarios where domain boundaries arise from heterogeneous textual representations rather than uniformly
structured attributes. We partition the Abt-Buy dataset by \emph{brand}, again retaining only domains with at least 100 labeled samples.

\textbf{Scope and Generality.} It is worth noting that while our experiments focus on e-commerce product data, the EMAD formulation is applicable to other macro-domains. For example, in restaurant matching~\cite{mudgal2018deep}, different cuisine types (e.g., Indian, Italian, or vegan) may benefit from domain-specific models, while still sharing structural attributes such as names, descriptions, and locations. This shared structure suggests that distribution-aware cross-category sampling--analogous to BEACON’s use of product categories--could also be effective in such settings. At the same time, not all datasets admit a natural or useful domain partition, and some partitions may offer limited benefit due to weak domain boundaries. For instance, defining domains by \emph{price range} yielded unimpressive results in our offline experiments. Nevertheless, we view e-commerce product data as a representative and practically relevant testbed, and contend that the principles underlying EMAD could extend to other domains where meaningful substructure exists.

\subsection{Implementation Details} \label{sub:implementation_details}

All experiments were conducted on NVIDIA A100 (80GB) GPUs, each part of a high-performance computing cluster running Linux and Python~3.7.

For each experiment, we construct a unique training pool and domain configuration, with total training pools ranging from approximately 5k to 20k candidate pairs and the number of domains ranging from 6 to 13. In our main results (Section~\ref{sec:main_results}), models are evaluated across ascending training budgets (e.g., 1k, 2k, ..., 10k)

A training-sample budget represents a practical constraint imposed by an end user of an EM system--reflecting the maximum number of labeled candidate pairs available for fine-tuning. To simulate this setting, we evaluate across a diverse range of budgets that represent users with varying levels of labeling capacity, thereby examining how model performance scales with budget size.


For all reported experiments, we fine-tune \texttt{RoBERTa}~\cite{liu2019roberta} as the PLM backbone, generating 768-dimensional embeddings. We adopt the default hyperparameters from DITTO~\cite{li2020ditto}, including its \emph{summarize} optimization for efficient fine-tuning.

Following prior work in EM~\cite{ebraheem2017deeper, mudgal2018deep, li2020ditto}, we evaluate models using the \emph{F1 score}. We report both \textbf{macro-average F1} (averaged equally across domains) and \textbf{weighted-average F1} (weighted by test set size per domain). The macro F1 emphasizes performance fairness across domains, while the weighted F1 captures overall matching accuracy across all samples. Formally, for F1 scores $F1_i$ computed on the test set $t_i$ of each domain $i \in \mathcal{D}$, we define:
\begin{equation}
\bar{F1}_{macro} = \frac{1}{|\mathcal{D}|} \sum_{i\in\mathcal{D}} F1_i, 
\quad
\bar{F1}_{weighted} = \frac{\sum_{i\in\mathcal{D}} F1_i \cdot |t_i|} {\sum_{i\in\mathcal{D}} |t_i|},
\end{equation}
where $\bar{F1}_{macro}$ is the unweighted domain-average F1, and $\bar{F1}_{weighted}$ weights each domain’s F1 by the number of test pairs $|t_i|$.

\subsection{Baselines} \label{sub:baselines}


To benchmark our proposed methods, we include a diverse set of baseline models:
\begin{itemize}
    \item \textbf{SPEC} and \textbf{GEN}: Two 
    simple baselines for EMAD (see Section~\ref{sec:basic_methods}). SPEC trains with only in-domain data, while GEN samples uniformly across all domains without distinction.\\
    \item \textbf{MFSN}~\cite{sun2024mfsn}: The state-of-the-art \emph{domain adaptation} framework for EM. MFSN (Matching Feature Separation Network) explicitly models both \emph{domain-common} and \emph{domain-private} matching features to align feature distributions between source and target domains. To enable a fair comparison within the EMAD setting, we adapt MFSN by treating the labeled in-domain data as the target domain and a budgeted random subset of labeled out-of-domain data as the source domain. Under this adaptation, MFSN is trained using \emph{both} in-domain and out-of-domain labeled data, allowing it to perform cross-domain adaptation under a fixed annotation budget within our experimental pipeline.\\ 
    
    \item \textbf{Battleship~\cite{genossar2023battleship}}: A state-of-the-art active learning framework for EM that selects samples using graph-based centrality and uncertainty. We adapt Battleship to the EMAD setting by first training it on in-domain data, and then using the resulting model to actively select a budgeted set of out-of-domain samples to add to the in-domain training pool. \\
    
    \item \textbf{PromptEM~\cite{promptEM2022}}: A state-of-the-art low-resource EM baseline that leverages PLM-based prompt tuning and pseudo-label generation to improve performance when annotated data are scarce. We train PromptEM using in-domain labels only in order to isolate the effectiveness of modern low-resource EM techniques without incorporating out-of-domain sample selection. \\

    \item \textbf{LLaMA-3.1}~\cite{dubey2024llama}: An open-source LLM with 8 billion parameters that achieves state-of-the-art performance among publicly available models of comparable scale. We use the \texttt{LLaMA-3.1-8B} model, adapting the "general-complex" prompt structure and zero-shot inference approach described by Peeters et. al.~\cite{peeters2023llm} to perform EM. This baseline evaluates the general reasoning and semantic understanding capabilities of modern LLMs on EM tasks.\\

    \item \textbf{Jellyfish}~\cite{zhang2023jellyfish}: A fine-tuned version of \texttt{LLaMA-3.1-8B} designed for structured data processing tasks, including EM. Jellyfish extends LLaMA through supervised fine-tuning on multiple data-centric tasks such as table understanding, schema matching, and EM. Notably, its fine-tuning corpus includes several EM benchmarks--such as the product-based Amazon–Google dataset~\cite{kopcke2010evaluation}--making it a particularly relevant and competitive baseline for our EMAD evaluation. Jellyfish is evaluated without additional training using its recommended EM prompting template.
\end{itemize}

Note that LLM-based methods do not have an explicit training budget since the size of their training data is not publicly available. Their performance is reported as constant across budgets for comparison.

\section{Results and Analysis} \label{sec:experiments}
In this section, we present the results of evaluating the BEACON framework across our domain-partitioned datasets. Section~\ref{sec:main_results} reports the main experimental results. Section~\ref{sec:ensemble_ablation} analyzes how BEACON functions as an ensemble of models. Section~\ref{sec:category_ablation} examines per-category performance under varying budget constraints. Section~\ref{sec:unbudgeted} compares the performance of BEACON against baseline methods without a budget.

\begin{table*}[t]
  \centering
  \caption{Macro F1 Performance for 50\% CC WDC Datasets }
  \label{tab:50cc_results}
  \footnotesize
  \makebox[\linewidth][c]{%
    \begin{subtable}[t]{0.31\linewidth}
      \centering
      \caption{0\% Unseen}
      \resizebox{\linewidth}{!}{\begin{tabular}{lccc|c}
\toprule
Method & 5.0k & 10.0k & Mean & SD \\
\midrule
SPEC & \underline{0.763} & \underline{0.770} & \underline{0.743} & 0.045 \\
GEN & 0.724 & 0.738 & 0.706 & 0.043 \\
\textbf{BEACON (ours)} & \textbf{0.778} & \textbf{0.790} & \textbf{0.768} & 0.034 \\
MFSN~\cite{sun2024mfsn} & 0.710 & 0.724 & 0.712 & 0.013 \\
\revGeneral{BATTLESHIP~\cite{genossar2023battleship}} & \revGeneral{0.699} & \revGeneral{0.697} & \revGeneral{0.676} & \revGeneral{0.045} \\
\revA{PROMPTEM~\cite{promptEM2022}} & \revA{0.610} & \revA{0.613} & \revA{0.609} & \revA{0.008} \\
LLAMA~\cite{dubey2024llama} & 0.653 & 0.653 & 0.653 & 0.000 \\
JELLYFISH~\cite{zhang2023jellyfish} & 0.625 & 0.625 & 0.625 & 0.000 \\
\bottomrule
\end{tabular}
}
    \end{subtable}\hspace{0.02\linewidth}%
    \begin{subtable}[t]{0.31\linewidth}
      \centering
      \caption{50\% Unseen}
      \resizebox{\linewidth}{!}{\begin{tabular}{lccc|c}
\toprule
Method & 5.0k & 10.0k & Mean & SD \\
\midrule
SPEC & 0.655 & 0.688 & 0.660 & 0.034 \\
GEN & 0.660 & \underline{0.709} & 0.656 & 0.057 \\
\textbf{BEACON (ours)} & \textbf{0.758} & \textbf{0.769} & \textbf{0.752} & 0.025 \\
MFSN~\cite{sun2024mfsn} & 0.658 & 0.631 & 0.637 & 0.020 \\
\revGeneral{BATTLESHIP~\cite{genossar2023battleship}} & \revGeneral{0.661} & \revGeneral{0.674} & \revGeneral{0.643} & \revGeneral{0.041} \\
\revA{PROMPTEM~\cite{promptEM2022}} & \revA{0.583} & \revA{0.588} & \revA{0.584} & \revA{0.005} \\
LLAMA~\cite{dubey2024llama} & 0.659 & 0.659 & 0.659 & 0.000 \\
JELLYFISH~\cite{zhang2023jellyfish} & \underline{0.692} & 0.692 & \underline{0.692} & 0.000 \\
\bottomrule
\end{tabular}}
    \end{subtable}\hspace{0.02\linewidth}%
    \begin{subtable}[t]{0.31\linewidth}
      \centering
      \caption{100\% Unseen}
      \resizebox{\linewidth}{!}{\begin{tabular}{lccc|c}
\toprule
Method & 5.0k & 10.0k & Mean & SD\\
\midrule
SPEC & 0.584 & 0.580 & 0.573 & 0.020\\
GEN & 0.607 & 0.622 & 0.602 & 0.033\\
\textbf{BEACON (ours)} & \textbf{0.652} & \textbf{0.664} & \textbf{0.645} & 0.024 \\
MFSN~\cite{sun2024mfsn} & 0.553 & 0.541 & 0.526 & 0.022\\
\revGeneral{BATTLESHIP~\cite{genossar2023battleship}} & \revGeneral{0.596} & \revGeneral{0.591} & \revGeneral{0.577} & \revGeneral{0.034} \\
\revA{PROMPTEM~\cite{promptEM2022}} & \revA{0.529} & \revA{0.533} & \revA{0.527} & \revA{0.006} \\
LLAMA~\cite{dubey2024llama} & \underline{0.631} & \underline{0.631} & \underline{0.631} & 0.000\\
JELLYFISH~\cite{zhang2023jellyfish} & 0.611 & 0.611 & 0.611 & 0.000\\
\bottomrule
\end{tabular}}
    \end{subtable}%
  }
\end{table*}
\begin{table*}[t]
  \centering
  \caption{Weighted F1 Performance for 50\% CC WDC Datasets}
  \label{tab:50cc_results_w}
  \footnotesize
  \makebox[\linewidth][c]{%
    \begin{subtable}[t]{0.31\linewidth}
      \centering
      \caption{0\% Unseen}
      \resizebox{\linewidth}{!}{\begin{tabular}{lccc|c}
\toprule
Method & 5.0k & 10.0k & Mean & SD \\
\midrule
SPEC & \textbf{0.777} & \underline{0.782} & \underline{0.746} & 0.061 \\
GEN & 0.728 & 0.761 & 0.717 & 0.055 \\
\textbf{BEACON (ours)} & \textbf{0.777} & \textbf{0.783} & \textbf{0.763} & 0.039 \\
MFSN~\cite{sun2024mfsn} & \underline{0.746} & 0.750 & 0.733 & 0.029 \\
\revGeneral{BATTLESHIP~\cite{genossar2023battleship}} & \revGeneral{0.678} & \revGeneral{0.700} & \revGeneral{0.664} & \revGeneral{0.047} \\
\revA{PROMPTEM~\cite{promptEM2022}} & \revA{0.718} & \revA{0.724} & \revA{0.712} & \revA{0.019} \\
LLAMA~\cite{dubey2024llama} & 0.575 & 0.575 & 0.575 & 0.000 \\
JELLYFISH~\cite{zhang2023jellyfish} & 0.676 & 0.676 & 0.676 & 0.000 \\
\bottomrule
\end{tabular}}
    \end{subtable}\hspace{0.02\linewidth}%
    \begin{subtable}[t]{0.31\linewidth}
      \centering
      \caption{50\% Unseen}
      \resizebox{\linewidth}{!}{\begin{tabular}{lccc|c}
\toprule
Method & 5.0k & 10.0k & Mean & SD \\
\midrule
SPEC & \underline{0.710} & \underline{0.731} & \underline{0.697} & 0.048 \\
GEN & 0.694 & 0.729 & 0.682 & 0.064 \\
\textbf{BEACON (ours)} & \textbf{0.758} & \textbf{0.763} & \textbf{0.739} & 0.035 \\
MFSN~\cite{sun2024mfsn} & 0.681 & 0.680 & 0.668 & 0.022 \\
\revGeneral{BATTLESHIP~\cite{genossar2023battleship}} & \revGeneral{0.647} & \revGeneral{0.673} & \revGeneral{0.634} & \revGeneral{0.047} \\
\revA{PROMPTEM~\cite{promptEM2022}} & \revA{0.669} & \revA{0.683} & \revA{0.669} & \revA{0.014} \\
LLAMA~\cite{dubey2024llama} & 0.595 & 0.595 & 0.595 & 0.000 \\
JELLYFISH~\cite{zhang2023jellyfish} & 0.691 & 0.691 & 0.691 & 0.000 \\
\bottomrule
\end{tabular}}
    \end{subtable}\hspace{0.02\linewidth}%
    \begin{subtable}[t]{0.31\linewidth}
      \centering
      \caption{100\% Unseen}
      \resizebox{\linewidth}{!}{\begin{tabular}{lccc|c}
\toprule
Method & 5.0k & 10.0k & Mean & SD \\
\midrule
SPEC & 0.644 & 0.646 & 0.631 & 0.032 \\
GEN & \underline{0.668} & \underline{0.685} & \underline{0.657} & 0.044 \\
\textbf{BEACON (ours)} & \textbf{0.701} & \textbf{0.717} & \textbf{0.697} & 0.021 \\
MFSN~\cite{sun2024mfsn} & 0.590 & 0.588 & 0.579 & 0.017 \\
\revGeneral{BATTLESHIP~\cite{genossar2023battleship}} & \revGeneral{0.632} & \revGeneral{0.641} & \revGeneral{0.617} & \revGeneral{0.038} \\
\revA{PROMPTEM~\cite{promptEM2022}} & \revA{0.597} & \revA{0.599} & \revA{0.592} & \revA{0.008} \\
LLAMA~\cite{dubey2024llama} & 0.625 & 0.625 & 0.625 & 0.000 \\
JELLYFISH~\cite{zhang2023jellyfish} & 0.649 & 0.649 & 0.649 & 0.000 \\
\bottomrule
\end{tabular}}
    \end{subtable}%
  }
\end{table*}

\subsection{Main Results} \label{sec:main_results}
We present the results of our extensive experimental evaluation in Tables~\ref{tab:50cc_results}, \ref{tab:50cc_results_w}, and \ref{tab:other_main_results}. While we evaluate model performance across multiple annotation budgets for each dataset (e.g., from 1k to 10k), we report results for two representative budgets (5k and 10k) for brevity.

\textbf{Performance on the WDC Multi-Dimensional Benchmark}. We conduct experiments on the WDC datasets with 50\% corner cases across ten annotation budgets ranging from 1k to 10k samples. Macro-averaged results are reported in Table~\ref{tab:50cc_results}, with corresponding weighted results shown in Table~\ref{tab:50cc_results_w}. In the macro setting, BEACON achieves the highest F1 scores at both $\beta=5$k and $\beta=10$k, as well as the highest mean performance across all ten budgets. Specifically, BEACON improves macro-average F1 by \textbf{2.5\%}, \textbf{6.0\%}, and \textbf{1.4\%} on the seen, half-seen, and unseen test datasets, respectively--yielding an average improvement of \textbf{3.3\%} over the next-best method. We attribute this gain to BEACON’s ability to construct budget-constrained training sets that are both representative and well-aligned in embedding space.

In the weighted setting, BEACON again outperforms all baselines, improving weighted F1 by \textbf{2.3\%}, \textbf{4.2\%}, and \textbf{4.0\%} for the seen, half-seen, and unseen datasets, respectively--an average improvement of \textbf{3.5\%}. The generally higher weighted F1 values suggest that domains with more training data benefit more directly from effective sample selection.

We observe several additional trends. First, \textbf{SPEC} performs best when the test set contains no unseen entities. This behavior is expected, as SPEC is trained exclusively on in-domain data and therefore has direct exposure to the distribution of entities present in the test set. Conversely, the \textbf{GEN} model performs better relative to \textbf{SPEC} when the test set consists entirely of unseen entities. By sampling broadly across all domains, GEN learns more general representations that transfer better to unfamiliar domains, yielding stronger performance in fully unseen settings.

Interestingly, the performance of the LLM-based models (LLaMA\hyp{}3.1 and Jellyfish) varies across the different unseen-entity configurations. Since these models are not fine-tuned on WDC data, the notion of “seen” versus “unseen” entities does not directly apply. Instead, the observed variation likely reflects differences in domain composition and difficulty across datasets--for example, some datasets may contain more linguistically complex or attribute-rich samples that better align with the LLMs’ pretraining distributions. 

PromptEM achieves lower F1 scores than the other methods, particularly in the macro setting. This behavior reflects its design: in our evaluation, PromptEM is trained using in-domain labels only and does not incorporate out-of-domain samples. As a result, its performance highlights that data-scarce domains often benefit substantially from leveraging cross-domain information, and that low-resource techniques alone may be insufficient to consistently optimize performance across a diverse set of domains. We further observe that BEACON consistently outperforms the active learning baseline Battleship. Although Battleship has demonstrated state-of-the-art performance in traditional single-domain active learning settings for EM, its selection strategy does not explicitly account for domain structure. This suggests that methods designed for single-domain active learning do not directly transfer to budgeted, cross-domain EM scenarios without explicit domain-aware adaptations, as incorporated in BEACON.

\begin{table*}[t]
  \centering
  \caption{Weighted F1 Performance on Additional EMAD Experiments}
  \label{tab:other_main_results}
  \footnotesize
  \makebox[\linewidth][c]{%
    \begin{subtable}[t]{0.31\linewidth}
      \centering
      \caption{WDC Product Corpus}
      \resizebox{\linewidth}{!}{\begin{tabular}{lccc|c}
\toprule
\revA{Method} & \revA{10.0k} & \revA{20.0k} & \revA{Mean} & \revA{SD} \\
\midrule
\revA{SPEC} & \revA{\underline{0.871}} & \revA{0.868} & \revA{\underline{0.861}} & \revA{0.017} \\
\revA{GEN} & \revA{0.853} & \revA{\underline{0.883}} & \revA{0.850} & \revA{0.029} \\
\revA{\textbf{BEACON (ours)}} & \revA{\textbf{0.877}} & \revA{\textbf{0.886}} & \revA{\textbf{0.873}} & \revA{0.022} \\
\revA{MFSN~\cite{sun2024mfsn}} & \revA{0.870} & \revA{0.875} & \revA{0.857} & \revA{0.026} \\
\revA{BATTLESHIP~\cite{genossar2023battleship}} & \revA{0.756} & \revA{0.755} & \revA{0.748} & \revA{0.020} \\
\revA{PROMPTEM~\cite{promptEM2022}} & \revA{0.820} & \revA{0.815} & \revA{0.814} & \revA{0.014} \\
\revA{LLAMA~\cite{dubey2024llama}} & \revA{0.767} & \revA{0.767} & \revA{0.767} & \revA{0.000} \\
\revA{JELLYFISH~\cite{zhang2023jellyfish}} & \revA{0.837} & \revA{0.837} & \revA{0.837} & \revA{0.000} \\
\bottomrule
\end{tabular}
}
      \label{tab:wdc4_results}
    \end{subtable}\hspace{0.02\linewidth}%
    \begin{subtable}[t]{0.31\linewidth}
      \centering
      \caption{Abt-Buy}
      \resizebox{\linewidth}{!}{\begin{tabular}{lccc|c}
\toprule
\revA{Method} & \revA{2.0k} & \revA{3.5k} & \revA{Mean} & \revA{SD} \\
\midrule
\revA{SPEC} & \revA{\underline{0.836}} & \revA{0.839} & \revA{\underline{0.830}} & \revA{0.014} \\
\revA{GEN} & \revA{0.772} & \revA{\underline{0.870}} & \revA{0.796} & \revA{0.087} \\
\revA{\textbf{BEACON (ours)}} & \revA{\textbf{0.839}} & \revA{\textbf{0.885}} & \revA{\textbf{0.858}} & \revA{0.024} \\
\revA{MFSN~\cite{sun2024mfsn}} & \revA{0.714} & \revA{0.696} & \revA{0.708} & \revA{0.006} \\
\revA{PROMPTEM~\cite{promptEM2022}} & \revA{0.692} & \revA{0.704} & \revA{0.695} & \revA{0.008} \\
\revA{BATTLESHIP~\cite{genossar2023battleship}} & \revA{0.462} & \revA{0.559} & \revA{0.487} & \revA{0.073} \\
\revA{LLAMA~\cite{dubey2024llama}} & \revA{0.366} & \revA{0.366} & \revA{0.366} & \revA{0.000} \\
\revA{JELLYFISH~\cite{zhang2023jellyfish}} & \revA{0.797} & \revA{0.797} & \revA{0.797} & \revA{0.000} \\
\bottomrule
\end{tabular}
}
      \label{tab:abtbuy}
    \end{subtable}\hspace{0.02\linewidth}%
    \begin{subtable}[t]{0.31\linewidth}
      \centering
      \caption{Ensemble Ablation (WDC)}
      \resizebox{\linewidth}{!}{\begin{tabular}{lccc|c}
\toprule
Method & 5.0k & 10.0k & Mean & SD \\
\midrule
SPEC & 0.710 & 0.731 & 0.697 & 0.048 \\
GEN & 0.694 & 0.729 & 0.682 & 0.064 \\
NN & 0.722 & 0.741 & 0.705 & 0.059 \\
TVDF & 0.739 & 0.745 & 0.717 & 0.046 \\
KCG & 0.718 & 0.747 & \underline{0.719} & 0.032 \\
KCG-TVDF & \textbf{0.758} & \underline{0.763} & \textbf{0.739} & 0.035 \\
ALL & \underline{0.747} & \textbf{0.764} & \textbf{0.739} & 0.034 \\
\bottomrule
\end{tabular}}
      \label{tab:ensemble_models}
    \end{subtable}%
  }
\end{table*}

\textbf{Experiments with Larger Annotation Pools}. Beyond the nine multi-dimensional WDC datasets, BEACON also achieves superior performance on the WDC Products Corpus. For this dataset, we run 10 budgets ranging from $\beta=2$k to $\beta=20$k in increments of 2k samples to account for the larger domains. The corresponding results are reported in Table~\ref{tab:wdc4_results}. These findings suggest that our methods are robust to variation in both the size and distribution of annotation pools across datasets. Moreover, they demonstrate that distribution-aware out-of-domain selection can improve performance even when the number of in-domain samples is already sufficient for fine-tuning. We therefore conclude that BEACON provides a robust solution for EM under tight data budgets and diverse domain conditions.

\textbf{Experiments with Diverse Schema Structures.} To evaluate EMAD on Abt-Buy, we experiment with five annotation budgets ranging from $\beta=1.5$k to $\beta=3.5$k in increments of 500 samples, with results shown in Table~\ref{tab:abtbuy}. BEACON outperforms all baselines on this dataset at the 2k and 3.5k budget levels, achieving the highest mean F1 score across all evaluated budgets. Notably, we observe substantially less performance variation across budgets on Abt-Buy than on the other datasets (i.e., lower standard deviation). We hypothesize that this effect stems primarily from the choice of domain partition: grouping products by \emph{brand} may induce greater homogeneity within domains than partitioning by \emph{product category}, as brands often focus on a narrower range of products with more consistent matching characteristics. Nevertheless, most approaches still benefit from additional out-of-domain samples, with performance generally improving from the 2k to the 3.5k budget. Another point of interest is the substantial performance improvement obtained through fine-tuning for LLM-based methods. While LLaMA achieves an F1 score of 0.366 in a zero-shot setting, fine-tuning via Jellyfish more than doubles performance, reaching an F1 score of 0.797. This suggests that fine-tuning is particularly effective for predominantly textual or unstructured EM tasks, further motivating the need for principled data selection strategies to guide fine-tuning under limited budgets. Overall, these results suggest that BEACON remains effective on textual EM datasets and further highlight its robustness and practical utility across diverse EM settings.

\subsection {Ablation Study} \label{sec:ensemble_ablation}

To isolate the effect of ensembling, we fix the evaluation dataset to be the 50\% corner case, 50\% unseen variant of WDC. 
We also report only weighted F1 to focus on raw predictive performance, without concern for domain size balance. All other experimental settings mirror those described in Section~\ref{sec:main_results}.

We evaluate a range of model combinations drawn from the methods introduced in Sections~\ref{sec:basic_methods} and ~\ref{sec:dfm}. Table~\ref{tab:ensemble_models} summarizes the results. While we experimented with all possible combinations, we report only a representative subset to summarize the key outcomes. Specifically, Table~\ref{tab:ensemble_models} includes single-model results for each method, a two-model ensemble (KCG-TVDF), and a five-model ensemble that aggregates all proposed methods. Overall, ensemble models outperform individual ones, with \textbf{KCG-TVDF} and \textbf{KCG-TVDF-NN-SPEC-GEN} achieving the best or second-best F1 scores across all tested budgets. This confirms that ensembling multiple models has a consistently positive effect on EM performance.

Notably, many ensembles--including those shown in Table~\ref{tab:ensemble_models}--achieve nearly identical mean F1 scores across budgets, suggesting similar levels of overall predictive power. Among them, \textbf{KCG-TVDF} stands out as the most practical choice due to its simplicity and robustness: it combines only two models yet reaches top-tier performance with minimal ensemble complexity. We therefore adopt \textbf{KCG-TVDF} as the core ensemble underlying BEACON.

\subsection{Performance of Individual Domains} \label{sec:category_ablation}

We now turn to a per-domain analysis using the WDC dataset with 50\% corner cases and 50\% unseen entities. Specifically, we examine individual domain performance and analyze how our distribution-aware methods allocate out-of-domain samples when augmenting a single target domain (Computers).

To evaluate the matching performance of individual domains, we examine the 12 product categories that contain at least 100 total samples. The experimental setup is identical to that of Section~\ref{sec:main_results}, except that we report performance for each individual product category rather than macro or weighted averages. Results are summarized in Table~\ref{tab:category}, showing F1 scores for $\beta=5$k and $\beta=10$k budgets and their mean across budgets. To contextualize BEACON’s performance, we also report SPEC--the second-best baseline in this setting.

Interestingly, EM performance does not correlate directly with the number of samples 
in a given domain. For example, the \emph{Cellphones} category achieves an average F1 score of 0.980 across ten budgets despite having only 197 total samples. In contrast, \emph{Computers}--the largest category--obtains a lower average F1 score (0.787).

This discrepancy likely stems from the inherent variation across domains. As discussed in Section~\ref{sub:prob_def}, differences in vocabulary, attribute composition, and matching criteria make some domains inherently more challenging for EM than others. More difficult domains may require substantially more labeled data to achieve comparable performance, naturally leading to larger variation across categories. In addition, smaller domains tend to be less stable: with limited  training and test data, their F1 scores become highly sensitive to the difficulty of individual samples. For instance, \emph{Cellphones} may perform well simply because its limited examples are relatively straightforward, whereas \emph{Tools}--another small domain--may contain a higher proportion of challenging or ambiguous pairs. Moreover, the limited samples in these smaller domains are less likely to be representative of their true underlying data distributions, further contributing to their observed performance instability. Consequently, most large domains exhibit more consistent F1 scores, typically in the 0.7–0.8 range.

\begin{figure}
    \centering
    \includegraphics[width=0.6\textwidth]{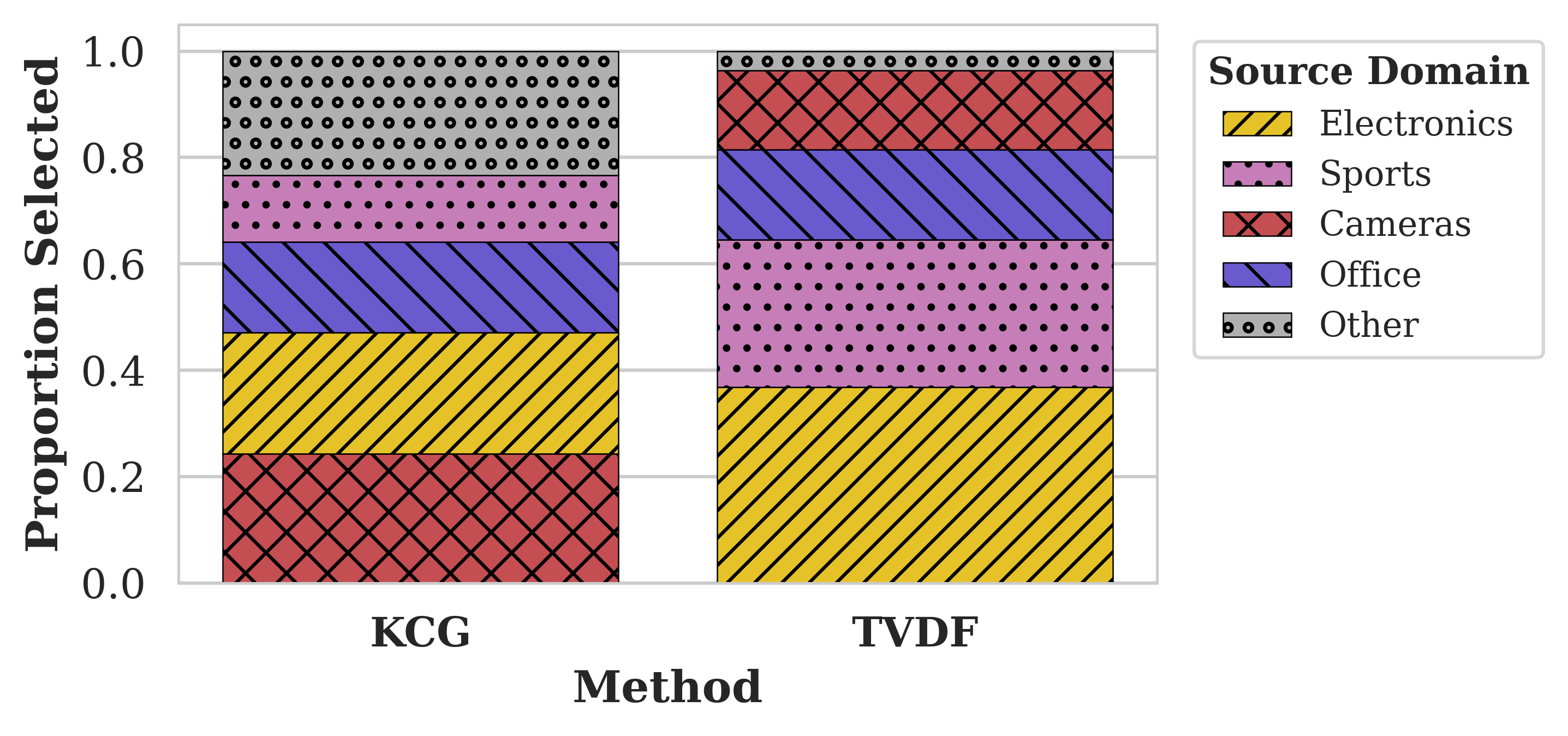}
    \caption{Analysis of out-of-domain sample selection for the \textit{Computers} target domain from WDC using the KCG and TVDF methods.}
    \label{fig:computers_selection_analysis}
\end{figure}

Figure~\ref{fig:computers_selection_analysis} illustrates how our distribution-aware methods select source domains when augmenting the \textit{Computers} target category. We conduct a controlled EMAD experiment with an 8k annotation budget and report the proportion of selected samples originating from each source domain for the KCG and TVDF selection strategies. As Figure~\ref{fig:computers_selection_analysis} suggests, KCG selects from a more diverse set of source domains, while TVDF concentrates its selections on fewer domains; however, both methods primarily favor semantically related domains (e.g., \textit{Cameras} and \textit{Electronics}).

\begin{table}[t]
\centering
\caption{WDC (50\%cc, 50\% unseen) Category Analysis.}
\label{tab:category}
\resizebox{0.55\textwidth}{!}{\begin{tabular}{lcccc}
\toprule
Category & 5.0k & 10.0k & Mean & SPEC \\
\midrule
Computers (6504) & 0.808 & 0.820 & \textbf{0.787} & \underline{0.778} \\
Electronics (2744) & 0.777 & 0.774 & \textbf{0.751} & \underline{0.736} \\
Office (1782) & 0.782 & 0.789 & \textbf{0.748} & \underline{0.706} \\
Sports (1705) & 0.646 & 0.594 & \textbf{0.610} & \underline{0.589} \\
Cameras (1580) & 0.755 & 0.762 & \textbf{0.755} & \underline{0.728} \\
Jewelry (460) & 0.645 & 0.656 & \textbf{0.645} & \underline{0.604} \\
Automotive (143) & 0.950 & 0.950 & \textbf{0.948} & \underline{0.891} \\
Clothing (199) & 0.626 & 0.742 & \underline{0.733} & \textbf{0.735} \\
Cellphones (139) & 0.989 & 1.000 & \textbf{0.980} & \underline{0.426} \\
Instruments (130) & 0.712 & 0.751 & \textbf{0.705} & \underline{0.675} \\
Tools (102) & 0.656 & 0.646 & \textbf{0.630} & \underline{0.546} \\
Home/Garden (70) & 0.745 & 0.744 & \textbf{0.735} & \underline{0.508} \\
\midrule
Weighted Average (15558) & 0.758 & 0.763 & \textbf{0.739} & \underline{0.697} \\
\bottomrule
\end{tabular}}
\end{table} 

\subsection{Unbudgeted Analysis} \label{sec:unbudgeted}

We now perform an analysis parallel to Section~\ref{sec:category_ablation}, but without imposing a training budget. This experiment compares BEACON against the GEN and SPEC models when each is allowed to use all available training data. For GEN, this corresponds to training a single model over the entire dataset of approximately 16{,}000 candidate pairs, while SPEC trains separate models using only domain-specific data for each category. In this context, GEN represents a fully pooled model that exploits all training data across domains, whereas SPEC serves as a purely domain-isolated baseline. 
We show the capabilities of BEACON (under a 10k budget) to improve the in-domain performance. The results are given in Table~\ref{tab:category_unbudgeted}.
For SPEC, the values in parentheses denote the number of samples in each domain, while for BEACON they indicate the absolute improvement in F1 score over SPEC. Overall, BEACON achieves higher F1 scores than both baselines despite using fewer samples than GEN. The largest gains appear in smaller domains such as \emph{Instruments}, where BEACON improves over the next-best baseline by 0.155.

In a few domains, GEN slightly outperforms BEACON--for example, in \emph{Clothing}, where GEN achieves an F1 score 0.10 higher. SPEC performs best in the largest domain, \emph{Computers}, though BEACON’s score is only 0.001 lower. On average, BEACON exceeds both unbudgeted baselines, reaching a mean F1 score of 0.763.

Another noteworthy observation from Table~\ref{tab:category_unbudgeted} is that both BEACON (at the 10k budget) and the unbudgeted GEN model achieve an F1 score of 1.0 on \emph{Cellphones}, whereas SPEC performs substantially worse, with an F1 of 0.393. Although this result partly reflects the volatility of smaller domains--given that \emph{Cellphones} is the fourth smallest domain--it also highlights the substantial benefit of incorporating out-of-domain samples. In this case, the inclusion of cross-domain data enables BEACON and GEN to achieve perfect category-level performance, underscoring the potential of out-of-domain sampling to stabilize small-domain training.

\begin{table}[t]
\centering
\caption{WDC (50\%cc, 50\% unseen) Unbudgeted 
Analysis.}
\label{tab:category_unbudgeted}
\resizebox{0.6\textwidth}{!}{%
\begin{tabular}{lcc|c}
\toprule
Category & SPEC & +BEACON ($\rightarrow$10k) & GEN ($\sim$16K)\\
\midrule
Computers & \textbf{0.821} (6504) & \underline{0.820} (-0.001) & 0.819 \\
Electronics & \underline{0.770} (2744) &  \textbf{0.774} (+0.004) & 0.737 \\
Office & 0.714 (1782) & \textbf{0.789} (+0.075) & \underline{0.786} \\
Sports & 0.317 (1705) & \underline{0.594} (+0.277) & \textbf{0.659} \\
Cameras &  \underline{0.702} (1580) & \textbf{0.762} (+0.060) & 0.698 \\
Jewelry & 0.293 (460) & \underline{0.656} (+0.363) & \textbf{0.683} \\
Clothing  & 0.308 (199) & \underline{0.742} (+0.434) & \textbf{0.842} \\
Automotive & 0.370 (143) & \textbf{0.950} (+0.580) & \underline{0.625} \\
Cellphones & \underline{0.393} (139) & \textbf{1.000} (+0.607) & \textbf{1.000} \\
Instruments & 0.330 (130) & \textbf{0.751} (+0.421) & \underline{0.596} \\
Tools & 0.438 (102) & \textbf{0.646} (+0.208) & \underline{0.625} \\
Home/Garden & 0.355 (70) & \textbf{0.744} (+0.389) & \underline{0.705} \\
\midrule
Weighted Avg  & 0.627 & \textbf{0.763} (+0.136) & \underline{0.754} \\
\bottomrule
\end{tabular}
}
\end{table} 

\section{Conclusion} \label{sec:conclusion}

This work introduces \emph{Budget-Aware Entity Matching Across Domains}, a new problem formulation aimed at training effective domain-specific matchers under strict sample budgets. To address this challenge, we propose \textbf{BEACON}, a distribution-aware sampling framework that selects out-of-domain samples to optimize domain-specific matching performance. We evaluate BEACON on multiple domain-partitioned datasets derived from several Entity Matching benchmarks, comparing it against strong baselines including low-resource, domain adaptation, and LLM-based methods. BEACON consistently outperforms these baselines, achieving higher macro and weighted F1 scores under equal or smaller budgets. While BEACON currently leverages a PLM backbone fine-tuned for EM, its principles extend to other embedding-based approaches. We plan to explore LLM backbones to assess cost–performance trade-offs and hybrid strategies that integrate embedding-based sampling with active learning to further enhance budget-aware EM.


\section*{Acknowledgments}
Large language models were used to assist with drafting portions of the manuscript, including text, tables, graphs, etc. This material is based upon work supported by the National Science Foundation under Grant NRT-HDR-2021871. Any opinions, findings, and conclusions or recommendations expressed in this material are those of the author(s) and do not necessarily reflect the views of the National Science Foundation.


\bibliographystyle{unsrt}
\bibliography{sample}


\newpage
\appendix
\section{Percentage of Corner Cases} \label{sec:cc_ablation}


We now examine the effect of varying the percentage of corner case samples in the WDC dataset. Corner cases are particularly difficult matching decisions--candidate pairs that are ambiguous or subtle enough that even human annotators may disagree on whether they refer to the same real-world entity. As noted by the authors of WDC~\cite{peeters2023wdc}, many EM methods in the literature, including DITTO~\cite{li2020ditto}, perform worse as the proportion of corner cases increases, and no known method is robust to them.
In this experiment, we fix the percentage of unseen entities in the test set to 50\% and vary the percentage of corner cases across three levels: 20\%, 50\%, and 80\%. We evaluate BEACON and the same set of baselines introduced in Section~\ref{sec:main_results}. Results are presented in Table~\ref{tab:cornercase_exp}.

Our findings align with those of Peeters et al~\cite{peeters2023wdc}: as the percentage of corner cases increases, overall model performance decreases--including for BEACON. Interestingly, however, the Jellyfish fine-tuned LLM performs slightly better on the 50\% corner case dataset than on the 20\% variant. This suggests that LLM-based models may be somewhat more resilient to corner cases than PLM-based models, possibly due to their larger parameter count or stronger generalization abilities.

We also observe small but consistent differences in the relative performance of BEACON across corner case settings. Specifically, BEACON improves over the next-best method by \textbf{4.4\%} on the 20\% corner case dataset, \textbf{4.2\%} on the 50\% dataset, and \textbf{1.9\%} on the 80\% dataset. This decreasing trend suggests that BEACON’s relative advantage may diminish as corner cases become more prevalent.

A plausible explanation is that a higher density of corner cases disrupts the structure of the embedding space used for sampling--reducing the separability of domain boundaries and weakening the effectiveness of embedding-based selection strategies. Similarly, the embedding-based EM method R-SupCon~\cite{peeters2022supervised} exhibits a comparable decline in effectiveness when trained and evaluated on datasets containing many corner cases~\cite{peeters2023wdc}.

\begin{table*}[t]
  \centering
  \caption{Weighted F1 Performance for 50\% Unseen WDC Datasets (Comparing Corner Cases)}
  \label{tab:cornercase_exp}
  \footnotesize
  \makebox[\linewidth][c]{%
    \begin{subtable}[t]{0.31\linewidth}
      \centering
      \caption{20\% Corner Cases}
      \resizebox{\linewidth}{!}{\begin{tabular}{lccc|c}
\toprule
Method & 5.0k & 10.0k & Mean & SD \\
\midrule
SPEC & \underline{0.765} & 0.767 & \underline{0.745} & 0.049 \\
GEN & 0.748 & \underline{0.780} & 0.739 & 0.052 \\
\textbf{BEACON (ours)} & \textbf{0.795} & \textbf{0.803} & \textbf{0.786} & 0.022 \\
MFSN~\cite{sun2024mfsn} & 0.706 & 0.733 & 0.718 & 0.016 \\
\revGeneral{BATTLESHIP~\cite{genossar2023battleship}} & \revGeneral{0.710} & \revGeneral{0.731} & \revGeneral{0.689} & \revGeneral{0.063} \\
\revA{PROMPTEM~\cite{promptEM2022}} & \revA{0.733} & \revA{0.739} & \revA{0.732} & \revA{0.011} \\
LLAMA~\cite{dubey2024llama} & 0.645 & 0.645 & 0.645 & 0.000 \\
JELLYFISH~\cite{zhang2023jellyfish} & 0.682 & 0.682 & 0.682 & 0.000 \\
\bottomrule
\end{tabular}}
    \end{subtable}\hspace{0.02\linewidth}%
    \begin{subtable}[t]{0.31\linewidth}
      \centering
      \caption{50\% Corner Cases}
      \resizebox{\linewidth}{!}{\begin{tabular}{lccc|c}
\toprule
Method & 5.0k & 10.0k & Mean & SD \\
\midrule
SPEC & \underline{0.710} & \underline{0.731} & \underline{0.697} & 0.048 \\
GEN & 0.694 & 0.729 & 0.682 & 0.064 \\
\textbf{BEACON (ours)} & \textbf{0.758} & \textbf{0.763} & \textbf{0.739} & 0.035 \\
MFSN~\cite{sun2024mfsn} & 0.681 & 0.680 & 0.668 & 0.022 \\
\revGeneral{BATTLESHIP~\cite{genossar2023battleship}} & \revGeneral{0.647} & \revGeneral{0.673} & \revGeneral{0.634} & \revGeneral{0.047} \\
\revA{PROMPTEM~\cite{promptEM2022}} & \revA{0.669} & \revA{0.683} & \revA{0.669} & \revA{0.014} \\
LLAMA~\cite{dubey2024llama} & 0.595 & 0.595 & 0.595 & 0.000 \\
JELLYFISH~\cite{zhang2023jellyfish} & 0.691 & 0.691 & 0.691 & 0.000 \\
\bottomrule
\end{tabular}}
    \end{subtable}\hspace{0.02\linewidth}%
    \begin{subtable}[t]{0.31\linewidth}
      \centering
      \caption{80\% Corner Cases}
      \resizebox{\linewidth}{!}{\begin{tabular}{lccc|c}
\toprule
Method & 5.0k & 10.0k & Mean & SD \\
\midrule
SPEC & \underline{0.703} & \underline{0.716} & \underline{0.682} & 0.054 \\
GEN & 0.682 & 0.710 & 0.651 & 0.078 \\
\textbf{BEACON (ours)} & \textbf{0.716} & \textbf{0.729} & \textbf{0.701} & 0.044 \\
MFSN~\cite{sun2024mfsn} & 0.668 & 0.673 & 0.674 & 0.007 \\
\revGeneral{BATTLESHIP~\cite{genossar2023battleship}} & \revGeneral{0.640} & \revGeneral{0.666} & \revGeneral{0.621} & \revGeneral{0.062} \\
\revA{PROMPTEM~\cite{promptEM2022}} & \revA{0.668} & \revA{0.667} & \revA{0.656} & \revA{0.017} \\
LLAMA~\cite{dubey2024llama} & 0.571 & 0.571 & 0.571 & 0.000 \\
JELLYFISH~\cite{zhang2023jellyfish} & 0.638 & 0.638 & 0.638 & 0.000 \\
\bottomrule
\end{tabular}}
    \end{subtable}%
  }
\end{table*}

\section{The Budget Distribution} \label{sec:budgets_ablation}

We now examine how varying the budget spread affects model performance. In Section~\ref{sec:main_results}, we selected a budget range of 1k–10k samples for WDC because most of its variants contain approximately 13k–15k training samples. This spread enables evaluation of out-of-domain selection strategies while (1) maintaining sufficient samples to fine-tune a PLM such as RoBERTa, and (2) avoiding the use of all available data, which would render the models equivalent or require oversampling. But what happens if we relax either condition (1) or (2)?  

\begin{figure*}[htbp]
  \centering
  \includegraphics[width=\linewidth]{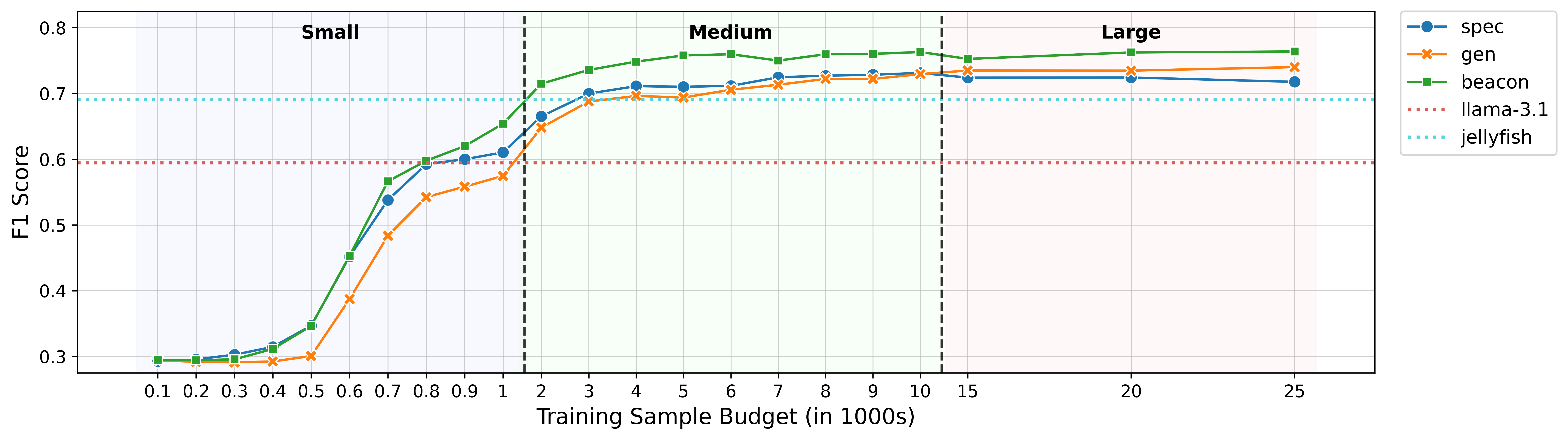}
  \caption{{A comparison of models from Table \ref{tab:50cc_results_w}} using small, medium, and large budget spreads.}
  \label{fig:budgets_ablation}
\end{figure*}

Figure~\ref{fig:budgets_ablation} presents results for three different budget spreads on the WDC 50\% corner case, 50\% seen dataset: our original spread, a smaller spread that relaxes condition (1), and a larger spread that relaxes condition (2). The smaller spread tests how well a PLM can be fine-tuned with very few samples, while the larger spread examines to what extent targeted oversampling can improve performance.  

As shown in Figure~\ref{fig:budgets_ablation}, all models exhibit lower overall matching performance under smaller budget spreads. BEACON achieves the highest mean F1 score in this range (0.444), followed closely by SPEC with a mean F1 of 0.435, and SPEC even slightly outperforming BEACON at some of the smallest budgets. This similarity is expected: when the available budget is very small, most models have little opportunity to draw from out-of-domain data, meaning that both SPEC and BEACON effectively rely on random selection from the same limited pool of in-domain samples. Consequently, as the budgets increase, BEACON benefits more from its targeted out-of-domain sampling, gradually separating itself from SPEC.

The large budget spread in Figure~\ref{fig:budgets_ablation} highlights the effect of targeted oversampling in the EMAD setting. Here, oversampling is guided by the selection criteria of each model, meaning that the “best” out-of-domain samples are reused rather than selected uniformly. For example, a KCG model using farthest-first selection will prioritize reusing the farthest point during oversampling instead of the closest.  

As shown in Figure~\ref{fig:budgets_ablation}, oversampling generally has a negligible effect on matching performance. Some models exhibit minor gains, as in the case of BEACON, which improves slightly between the 15k and 25k budgets. Others show small declines, such as SPEC, whose performance decreases marginally from 20k to 25k. Overall, the results suggest that model performance tends to stabilize once the budget enters the oversampling regime--indicating that oversampling offers limited additional benefit at large budgets, even though it can be beneficial for smaller domains and lower-budget settings.

\end{document}